\documentclass[12pt]{article}
\makeatletter
\renewcommand\paragraph{\@startsection{paragraph}{4}{\z@}%
	{-2.5ex\@plus -1ex \@minus -.25ex}%
	{1.25ex \@plus .25ex}%
	{\normalfont\normalsize\bfseries}}
	\makeatother
\usepackage{amsmath}
\usepackage[fleqn]{mathtools}
\usepackage{float}
\usepackage{slashed}
\usepackage[symbol]{footmisc}
\renewcommand{\thefootnote}{\fnsymbol{footnote}}
\usepackage{graphicx}
\usepackage{subfig}
\usepackage{hyperref}
\hypersetup{
        colorlinks=true,
    bookmarks=true,
    linkcolor=blue,
    filecolor=magenta,
    urlcolor=blue,
    citecolor=blue,
}
\def\d{{\rm d}}

\def\gev2{\hbox{GeV}^2}
\def\<{\langle}
\def\>{\rangle}
\def\lapp{\mathrel{\rlap{\raise.5ex\hbox{$<$}}
                    {\lower.5ex\hbox{$\sim$}}}}
\def\gapp{\mathrel{\rlap{\raise.5ex\hbox{$>$}}
                    {\lower.5ex\hbox{$\sim$}}}}
\def\tev{\, {\rm TeV}}

 \usepackage[normalem]{ulem}
\usepackage{color}

\def\lapp{\mathrel{\rlap{\raise.5ex\hbox{$<$}}
                    {\lower.5ex\hbox{$\sim$}}}}
\def\gapp{\mathrel{\rlap{\raise.5ex\hbox{$>$}}
                    {\lower.5ex\hbox{$\sim$}}}}

\begin{document}
\newpage

\begin{flushright}
\today
\end{flushright}

\begin{center}
{\Large \bf
Infrared finiteness of theories with bino-like \\ [0.3cm]
dark matter at finite temperature}\\ [0.5cm]
Pritam Sen\footnote[1]{pritamsen@imsc.res.in} and  D.
Indumathi\footnote[2]{indu@imsc.res.in} \\ [0.2cm]
The Institute of Mathematical Sciences, Chennai and
Homi Bhabha National Institute, Mumbai, India \\ [0.5cm]
Debajyoti Choudhury\footnote[3]{debajyoti.choudhury@gmail.com}\\[0.2cm]
Department of Physics and Astrophysics, University of Delhi,
Delhi 110 007, India \\ [1cm]

\end{center}

\renewcommand*{\thefootnote}{\arabic{footnote}}
\centerline{\underline{Abstract}} Models incorporating moderately heavy
dark matter (DM) typically need charged (scalar) fields to establish
admissible relic densities.  Since the DM freezes out at an early
epoch, thermal corrections to the cross sections can be important. In
a companion paper~\cite{paper1} we established that the infrared (IR)
divergences accruing from scalar-photon interactions cancel to all
orders in perturbation theory. The corresponding infrared finiteness of
thermal fermionic QED has already been established. Here, we study the IR
behaviour at finite temperatures, of a theory of dark matter interacting
with charged scalars and fermions, which potentially contains both both
linear and sub-leading logarithmic divergences. We prove that the theory
is IR-finite to all orders with the divergences cancelling when both
absorption and emission of photons from and into the heat bath are taken
into account. While 4-point interaction terms are known to be IR finite,
their inclusion leads to a neat exponentiation. The calculation follows
closely the technique used for the scalar finite temperature theory.

\vspace{0.3cm}

\noindent {\bf PACs}: 11.10.−z, 
11.10.Wx, 
11.15.−q, 
11.30.Pb, 
12.60.−i, 
95.35.+d 


\newpage

\section{Introduction}
\label{sec:intro}

Astrophysical and cosmological observations spanning over a multitude
of length scales, beginning with rotation curves, lensing, on to
galactic and cluster collisions, and finally the origin of large scale
structure in the Universe as also the power spectrum of the cosmic
microwave background radiation, all point to the existence of a
mysterious Dark Matter (DM) that overwhelms ordinary matter in the
Universe. And while  all the evidence so far has come
from the study of gravitational effects, simple modifications in the
theory of gravity have, so far, failed to account for all the
observations. The conundrum, apparently, can be resolved only by
postulating a particulate DM that, of necessity, must be immune to
strong interactions and, preferably, neutral\footnote{Although
milli-charged DM is still allowed, such models tend to be somewhat
contrived. The inclusion of this possibility, though, would not
change the central thesis of this paper, other than adding a layer
of complication to the calculations.}.

If we posit that the DM particle has no interactions at all with the
Standard Model (SM) particles, other than the gravitational, there
would be virtually no hope of ever observing it directly in a
controlled experiment. Consequently, it is assumed that the DM must
have sufficient interactions with at least some of the SM particles,
presumably with a strength comparable to weak interactions, or at
worst, a couple of order of magnitudes weaker. Such an assumption has
a further ramification.  A standard assumption in explaining the
evolution of the Universe is that all particles---whether those within
the SM, or the DM---were created during the (post-inflation) reheating
phase.  The subsequent number densities are supposed to have been
determined by the expansion of the Lemaitre-Friedmann-Robertson-Walker
Universe, augmented by a set of coupled Boltzmann equations that are
operative when the particles are in equilibrium. If the DM particle
$\chi$ does have such interactions, then it can stay in equilibrium
with the SM sector via interactions of the form,
\begin{equation}
  \chi + \overline{\chi} \leftrightarrow {\cal F}_{SM} + \overline {\cal F}_{SM} \ ,
  \qquad {\rm and} \qquad
  \chi  + {\cal F}_{SM}  \leftrightarrow \chi + {\cal F}_{SM} \ ,
\label{eq:dmcoll}
\end{equation}
where ${\cal F}_{SM}$ is a particle corresponding to some
arbitrary SM field. This equilibrium phase would last until the
interaction rate falls below the Hubble expansion rate.  With the
Universe cooling as it expands, the DM must fall out of equilibrium by
the time its mass exceeds the temperature. With large scale structure
formation liable to be destroyed in the presence of a dominantly hot
DM (i.e., one where the DM decoupled well before the temperature fell
down to $m_\chi$), the favourite scenario is that of a dominantly cold
DM (i.e., one which had become nonrelativistic at the decoupling
era). Interestingly, if the interactions governing Eq.~\ref{eq:dmcoll}
typically have a strength comparable to the weak interaction, then for
a wide range of masses, ${\cal O}(10) \hbox{ GeV} < m_\chi < {\cal
  O}(10^4)$ GeV, the relic density is of the required order.

Given the high precision to which the DM contribution to the energy
budget has been measured by the {\sc wmap}~\cite{WMAP} and,
subsequently, the {\sc planck}~\cite{planck} collaborations, an order
of magnitude estimation is no longer acceptable, and precise
predictions need to be made. Indeed, the measurements have imposed
rather severe constraints on several well-motivated models for the DM,
to the extent of ostensibly even ruling out some of them. Such
drastic conclusions, though, need to be treated with caution, for many
of the theoretical estimates have resulted from lowest-order
calculations alone. More importantly, the effect of non-zero
temperatures are rarely considered. Together, these effects can alter
the predictions to a significant degree.

Initial efforts to include thermal corrections to the relevant
processes were made in Ref.~\cite{Beneke}, wherein it was shown,
albeit to only the next-to-leading order (NLO), that infra-red
divergences (both soft and collinear) cancel out in processes
involving both charged scalars and charged fermions. However, no proof
to all orders exists so far and we provide one in this paper.  In an
earlier paper \cite{paper1}, referred to henceforth as Paper I, we had
discussed, in detail, the infra red (IR) behaviour, via factorisation
and resummation, of theories associated with charged scalar particles,
at both zero and finite temperature. In doing this, we proved the
IR finiteness of such theories to all orders of perturbation theory. The
IR finiteness of thermal fermionic QED has already been established to all
orders \cite{Indu}. In this second paper, we apply the results for thermal
fermionic and scalar theories to set out the analogous proof of the IR
finiteness of theories associated with bino-like dark matter (DM) models,
at non-zero temperatures. While IR finiteness is important to establish
the consistency of the theory, the inclusion of such corrections is also
crucial for an accurate calculation of the relic density of the DM.

Although it might seem that a bino-like DM candidate is a very
specific choice, it actually captures the essence of a wide class
of models. Whereas the MSSM spectrum would include, apart from the
SM particles, the entire gamut of their supersymmetric partners,
only a handful of them play a significant role in determining the
relic density. Apart from the DM candidate itself, these are some
charged scalars (typically, close to the DM in mass), and occasionally,
depending on the details of the supersymmetry breaking scheme, others
as well. And while the DM itself is a linear combination of the bino,
the neutral wino and the two neutral higgsinos, for a very large class
of supersymmetry breaking scenarios, the higgsino mass parameter $\mu$
is much larger than the soft terms ($M_{1,2}$) for the gauginos,
thereby suppressing the higgsino component to negligible levels. And
since wino-bino mixing is pivoted by $\mu$, a large value for the
latter also suppresses the wino-component in the DM. The assumption
of a bino-like DM further simplifies the calculations as we may safely
neglect additional diagrams, {\em e.g.}, with $s$-channel gauge bosons
or Higgs\footnote{For pure binos, such couplings arise only at one-loop
level, and are of little consequence.}.  It should also be appreciated
that no new infrared divergence structures would appear even on the
inclusion of such additional mediators\footnote{The only caveat to this
is presented by the diagrams involving the $W^\pm$, as photons could
also radiate off the latter. The structure of the ensuing IR divergences,
however, are quite analogous to those that we would encounter here, and
can be analysed similarly.}.  In other words, restricting ourselves to
the particular case of the bino does not represent the neglect of subtle
issues while allowing for considerable simplifications, both algebraic
and in bookkeeping.

The Lagrangian density relevant to this simplified scenario is given
by an extension of the Standard Model containing left handed fermion
doublets, ${f} = ({f}^0, {f}^-)^T$, with an additional scalar doublet,
$\phi = (\phi^+, \phi^0)^T$, namely the supersymmetric partners of
$f$, along with the $SU(2) \times U(1)$ singlet Majorana fermion
$\chi$ which is the dark matter candidate. We have,
\begin{align}
{\cal{L}} = & -\frac{1}{4} F_{\mu\nu} F^{\mu\nu} +
\overline{{f}} \left( i \slashed{D} - m_f \right) {{f}} + \frac{1}{2}
\overline{\chi} \left( i \slashed{D} - m_\chi \right) \chi \nonumber \\
  & + \left(D^\mu \phi \right)^\dagger \left(D_\mu \phi \right) -
  m_\phi^2 \phi^\dagger \phi + \left( \lambda \,\overline{\chi} P_L
  {{f}}^-
  \phi^+ + {\rm h.c.} \right)~.
\label{eq:L}
\end{align}
We assume that the bino is a TeV scale DM particle so that freeze-out
occurs {\em after} the electro-weak transition; hence, only
electromagnetic interactions are relevant for the IR finiteness at
these scales\footnote{Such an approximation is a very good one for
$m_\chi \lapp 2 \tev$. For $m_\chi \gapp 20 \tev$, again, one could
proceed in an entirely analogous fashion, replacing the photon by
the entire set of four electroweak gauge bosons. For an intermediate
mass bino, on the other hand, the analysis is rendered much more
complicated and is beyond the scope of this paper.}. In other words,
the $\chi$ interacts only with fermions and sfermions, ${{f}}$ and
$\phi$, and not with the photon. Thus, only the charged (s)fermion
interactions with $\chi$ are shown in Eq.~\ref{eq:L} since it is the
resummation of the radiative photon diagrams which are of interest here.

\begin{figure}[htp]
\begin{center}
\includegraphics[width=0.5\textwidth]{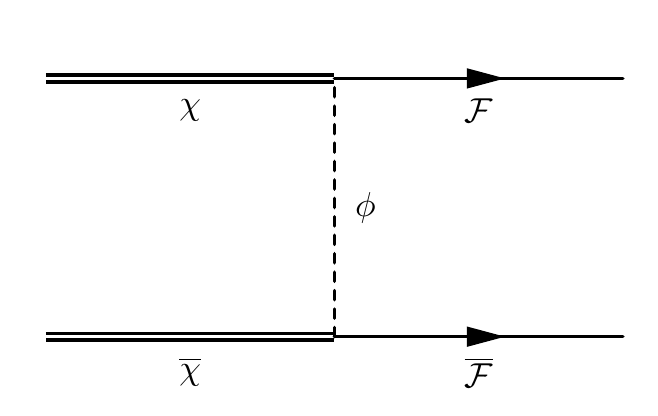}
\caption{\em A typical dark matter annihilation/scattering process.}
\label{fig:dm}
\end{center}
\end{figure}

The simplest process for DM annihilation (or, equivalently, DM
scattering off a SM particle), as driven by the Lagrangian of
Eq.~\ref{eq:L}, is illustrated in Fig.~\ref{fig:dm}. Higher order
electromagnetic corrections to such diagrams involve, apart from real
photon emissions from either $f$ or $\phi$, virtual photon
exchanges as well.

Such contributions can be calculated in a real time
formulation of the thermal field theory \cite{realtime}. In Ref.~\cite{Weldon}, the eikonal approximation was
used and the interaction of photons with a semi-classical current was
analysed within the framework of thermal field theory. In
Ref.~\cite{Donoghue}, 1-loop corrections to thermal QED were computed
for both fermions and scalars.  It was found that the finite
temperature mass shift for scalar QED when the temperature is less
than the scalar mass ($0 < T \ll m_s$) is identical to the
corresponding fermion case as is the IR divergent piece of the wave
function and vertex renormalisation constants. However, the
contribution to the plasma screening mass was twice
that for the fermion loop due to the
difference in the form of the thermal distributions (boson versus
fermion). In Ref.~\cite{dynRG}, dynamical renormalisation group
resummation of finite temperature IR divergences to 1-loop was done to
study relaxation as well as out-of-equilibrium damping in real time
for a scalar thermal field theory. This was applied to study IR
divergences in scalar QED (to the lowest order in the hard-thermal
loop (HTL) resummation).  It was found that the infrared divergences
in this theory are similar to those found in QED and in lowest order
in QCD. Results in thermal scalar QED have also been applied to study
Schwinger pair-production \cite{Schwinger}.

Here we use a similar approach to address the issue of IR finiteness
of the thermal field theory of dark matter, thereby combining and
extending the results of the earlier work on thermal fermions
\cite{Indu} with the finite temperature results for charged scalars as
discussed in Paper I \cite{paper1}.

In Section \ref{sec:tdark}, we set up the real-time formulation of the
thermal field theory corresponding to the Lagrangian of Eq.~\ref{eq:L} and
write down the propagators and vertex factors of the theory.  In Section
\ref{sec:irdm}, we set up the machinery to address the infrared (IR)
finiteness of such a theory of dark matter interacting with charged
fermions and scalars. In particular, we take the example of the process
$\chi {\cal{F}} \to \chi {\cal{F}}$ where the bino $\chi$ interacts with
fermions ${\cal{F}}$ via a bino-fermion-scalar interaction, and all higher
order contributions to this process. We outline in Section \ref{ssec:GY}
the approach of Grammer and Yennie \cite{GY} (henceforth referred to as
GY) which was used to address the IR finiteness of fermionic QED at zero
temperature through a detailed consideration of the process ${\cal{F}}(p)
\gamma^* (q) \to {\cal{F}}(p')$, which we use in our calculations, by
defining the so-called $K$ and $G$ photon insertions and the vertex $V$
that enables use of this approach (Section \ref{ssec:V}). In Section
\ref{ssec:kgamma}, we show that the $K$ photon insertions contain the IR
divergence and compute these contributions for all possible insertions
of the virtual $K$ photons. This section contains the main new results of
this paper. Since the calculation heavily depends on the results obtained
in earlier work \cite{paper1,Indu}, an overview of these results is also
presented here.

That the $G$ photon insertions are IR finite is shown in
Section \ref{ssec:ggamma}; this result straightforwardly follows from
the calculations done earlier \cite{paper1,Indu} and only highlights
are included. Emission and absorption of real photons from the heat bath
constitude the other set of corrections to the lowest order process of
interest. Since the real photon contribution is a straightforward extension
of earlier results \cite{paper1,Indu}, we simply outline the procedure,
and the main complications that arise before writing down the results
in Section \ref{ssec:realgamma}.

In Refs.~\cite{Weldon,Indu,Sourendu} it was shown that pure fermionic
thermal QED has both a {\em linear} divergence and a logarithmic
subdivergence in the infra red compared to the purely logarithmic
divergence encountered in the zero temperature theory, owing to the
nature of the thermal photon propagator. The same was true for the case
of thermal scalar QED as shown in Ref.~\cite{paper1} and is true here
as well. We factorise and exponentiate the divergences and show that
they cancel order by order between virtual and real photon contributions
(the latter include both emission and absorption terms) and hence prove
the IR finiteness of a thermal theory of dark matter to all orders.
We write down the total cross section to all orders in Section
\ref{ssec:cross} where we explicitly demonstrate the IR finiteness of
the cross section in the soft limit. Section \ref{sec:concl} contains
the discussions and conclusions. The appendices are used to set up the
Feynman rules (Appendix \ref{app:abino}) for thermal field theories
and to list some useful identities (Appendix \ref{app:fidentities})
that are used to factorise the $K$ photon contributions.

\section{Real-time formulation of the thermal field theory with dark
matter}
\label{sec:tdark}

We briefly review the real-time formulation \cite{realtime} of thermal
(scalar, fermion and photon) fields in equilibrium with a heat bath at
temperature $T$. The field theory of such a system is then a statistical
field theory with a thermal vacuum defined such that the ensemble average
of an operator can be written \cite{Rivers} as its expectation value
of time-ordered products in the thermal vacuum. In order to satisfy this
requirement, a special path is chosen for the integration in the complex
time plane; see Appendix~\ref{app:abino} for details. This results in
the fields satisfying the periodic boundary conditions,
\begin{equation*}
\varphi(t_0) = \pm \varphi(t_0 - i \beta)~,
\end{equation*}
where $\pm 1$ correspond to boson and fermion fields respectively.

There are two parts of the time path, $C_1$ and $C_2$, along
the real time axis and parallel to it (see details and figure in
Appendix~\ref{app:abino}); this gives rise to fields that ``live" on the
$C_1$ or $C_2$ line and hence leads to the well-known field-doubling,
so that fields are of type-1 (physical) or type-2 (ghosts), with
propagators acquiring a $2 \times 2$ matrix form. Only type-1 fields
can occur on external legs while fields of both types can occur on
internal legs, with the off-diagonal elements of the propagator allowing
for conversion of one type into another. The zero temperature part
of the propagator corresponds to the exchange of a virtual particle,
as usual, but the finite temperature part of the propagator represents
an on-shell contribution which measures the probability of emitting or
absorbing real particles from the medium. See Appendix~\ref{app:abino}
for detailed definitions of the scalar, fermion and photon field
propagators. Finally, the vertices, both 3-point and 4-point ones,
are modified in the thermal theory by extending the Lagrangian given in
Eq.~\ref{eq:L} to include both kinds of vertices \cite{thermal}. Details
are again in Appendix~\ref{app:abino}; we only note here that all the
fields at a given vertex {\em must be} of the same thermal type.

In particular, the photon propagator corresponding to a momentum $k$
can be expressed (in the Feynman gauge) as,
\begin{align}
i {\cal D}^{ab}_{\mu\nu} (k) & = - g_{\mu\nu} \, i D^{ab} (k)~, 
\label{eq:thermalD}
\end{align}
where the information on the field type is contained in $D^{ab} (k)$;
see Appendix~\ref{app:abino} for its definition. Note that the factor
$g_{\mu\nu}$ occurs in all components of the thermal photon propagator,
with the relevant part of the thermal photon propagator being
\begin{align}
i {\cal{D}}^{ab} (k) \sim &
	\left[ \frac{i}{k^2+i\epsilon} \delta^{ab} \pm  2 \pi \,
	\delta(k^2) \, N(\vert {k^0} \vert) \, 
	D^{ab}_{T} \right]~,
\label{eq:thermalprop}
\end{align}
where the first term corresponds to the $T=0$ contribution and the
second to the finite temperature part. While the fermionic number operator,
viz.,
\begin{equation} 
N_f(\vert {p^0} \vert)  \equiv
\frac{1}{\exp\{\vert p^0 \vert/T\}  + 1} \;
  \stackrel{p \to 0}{\longrightarrow} \; \frac{1}{2}~,
\end{equation}
is well-defined in the soft limit, the bosonic number operator in the
photon propagator contributes an additional power of ${k^0}$
in the denominator in the soft limit, since
\begin{equation} 
N(\vert k^0 \vert) \equiv 
	\frac{1}{\exp\{\vert k^0 \vert/T\}  - 1} 
	\; \stackrel{k \to 0}{\longrightarrow} \;
	\frac{T}{\vert k^0 \vert}~.
\end{equation}
Hence, it can be seen that the leading IR divergence in the finite
temperature part is linear rather than logarithmic as was the case at zero
temperature. Consequently, there is a residual logarithmic subdivergence
that must also be shown to cancel at finite temperatures, thus making
the generalisation to the thermal case non-trivial.

It was shown in two earlier papers \cite{paper1,Indu} that thermal field
theories of pure scalar and spinor electrodynamics are IR finite to all
orders in the theory. The object of this paper is to obtain an analogous
proof of IR finiteness for a thermal field theory of (charge-neutral)
dark matter interacting with both charged scalars and charged fermions
with these additional complications.

\section{The IR finiteness of the thermal field theory with dark matter}
\label{sec:irdm}

The companion paper, Paper I, established that a field theory of
charged scalars is IR finite to all orders both at $T=0$ and at finite
temperature. The corresponding all-order proof for fermionic QED is
already known \cite{Indu}. We now apply and extend these two results to
prove the IR finiteness to all orders of theories of dark matter at finite
temperature. The proof involves obtaining a neat factorisation and
exponentiation of the soft terms to all orders in the theory, with order
by order cancellation between the IR divergent contributions of virtual
and real photon corrections to the leading order contribution. 

\subsection{The Grammer Yennie approach}
\label{ssec:GY}

We use the approach of Grammer and Yennie \cite{GY}, as applied in the
earlier papers as well. We briefly outline their approach, which was used
to prove the IR finiteness of zero temperature fermionic QED. Their proof
technique involved starting at an $n^{\rm th}$ order correction (with
real or virtual photons) to the process $e(p) \gamma^*(q) \to e(p')$
(where $q$ is a hard momentum flowing in at the vertex $V$, and the
remaining $n$ photon momenta can be arbitrarily soft), and examining the
effect of adding an additional virtual or real $(n+1)^{\rm th}$ photon
with momentum $k$ to this graph, in all possible ways.

In particular, in order to separate out the IR divergent piece in a
virtual photon insertion, the corresponding photon propagator of the
newly inserted photon was written as the sum over so-called $K$- and
$G$-photon contributions, viz.,
\begin{align} \nonumber -i
\frac{g_{\mu{\nu}}}{k^2+i\epsilon} & = \frac{-i}{k^2+i\epsilon} \,
	\left\{\strut \left[\strut g_{\mu\nu} - b_k (p_f, p_i)\,k_\mu
	k_\nu \right] + \left[\strut b_k (p_f, p_i) k_\mu k_\nu \right]
	\right\}~, \nonumber \\
 & \equiv \frac{-i}{k^2+i\epsilon} \, \left\{\left[\strut  G\right] +
 \left[K\right] \right\}~,
\label{eq:gammaprop}
\end{align}
where $b_k$ (to be defined later below) is a function of $k$ as well
as the momenta, $p_f$, $p_i$, of the final and initial particles,
{\em i.e.}, just after (before) the final (initial) vertex, and
is defined so that all the IR divergences are collected into the
$K$-photon contributions. The key to this factorisation lies in
recognising that the insertion of a virtual $K$ photon vertex $\mu$ on
a fermion line is equivalent to the insertion $\gamma^\mu k_\mu \equiv
\slash{\!\!\!k}$  at the vertex $\mu$.  The contribution due to the
insertion can be expressed, courtesy generalised Feynman identities
(see Eq.~\ref{eq:fid} Appendix~\ref{app:fidentities}) as the {\em
difference} of two terms, leading to pair-wise cancellation between
different contributions. Ultimately the contribution of the insertion of
the $(n+1)^{\rm th}$ virtual $K$ photon is a single term, proportional
to the lower order matrix element, ${\cal{M}}_n$.

A similar separation of the polarisation sums for the insertion of real
photons into so-called $\widetilde{K}$ and $\widetilde{G}$ contributions
can be made. Again, GY showed that the insertion of a $\widetilde{K}$
real photon into an $n^{\rm th}$ order graph leads to a cross section
that is proportional to the lower order one. For both virtual and real
photon insertions, the $G$ ($\widetilde{G}$) photon contributions are IR
finite with the entire IR divergent contribution being contained in the
$K$ (and $\widetilde{K}$) contributions. Combining the virtual and real
photon contributions at every order, GY showed that the IR singularities
cancel between the real and virtual contributions.

It was shown in Ref.~\cite{Indu} that similar results hold for the
case of {\em thermal} fermionic QED as well, because of the presence
of the factor $g_{\mu\nu}$ in {\em all} components of the thermal
photon propagator as seen from Eq.~\ref{eq:thermalD}. Furthermore, it
was shown in Paper I~\cite{paper1} that similar
results hold for the case of thermal pure scalar QED as well since the
insertion of a virtual $K$ photon vertex, $\mu$, on a scalar line can
be simplified using an analogous set of generalised Feynman identities
which also yield differences of two terms (see Eq.~\ref{eq:sid} in
Appendix~\ref{app:fidentities}). A similar factorisation holds for
real photon emission as well. Ultimately, the IR singularities cancel
between the real and virtual contributions. A special feature of the
thermal case is that they also cancel {\em only} when both real photon
emission and absorption contributions are included, since the soft
photons can be both emitted into, and absorbed from, the heat bath.

\subsection{Choice of vertex $V$}
\label{ssec:V}

One of our goals is to establish that a similar factorisation is obtained
for the thermal dark matter Lagrangian arising from Eq.~\ref{eq:L} as
well, with a typical scattering process being $\chi \overline{\chi}
\to {\cal{F}} \overline{\cal{F}}$, or $\chi {\cal{F}} \to {\cal{F}}
\chi$. These are $2 \to 2$ scattering processes; hence, applying the GY
technique in order to obtain a factorisation of the IR divergent terms
requires an unambiguous separation of the $p$ and $p'$ legs with $V$
arbitrary, along with the clear understanding that the IR divergences
arise from the inclusion of soft photons. It is convenient to consider
the process $\chi(q+q') \,{\cal{F}}(p) \to {\cal{F}}(p') \, \chi(q')$
where the momenta of the particles are chosen so that the momentum
of the intermediate scalar for the lowest order process is $(p-q')$
so that $p'-p = q$, the hard momentum, as before.

\begin{figure}[bht]
\begin{center}
\includegraphics[width=0.6\textwidth]{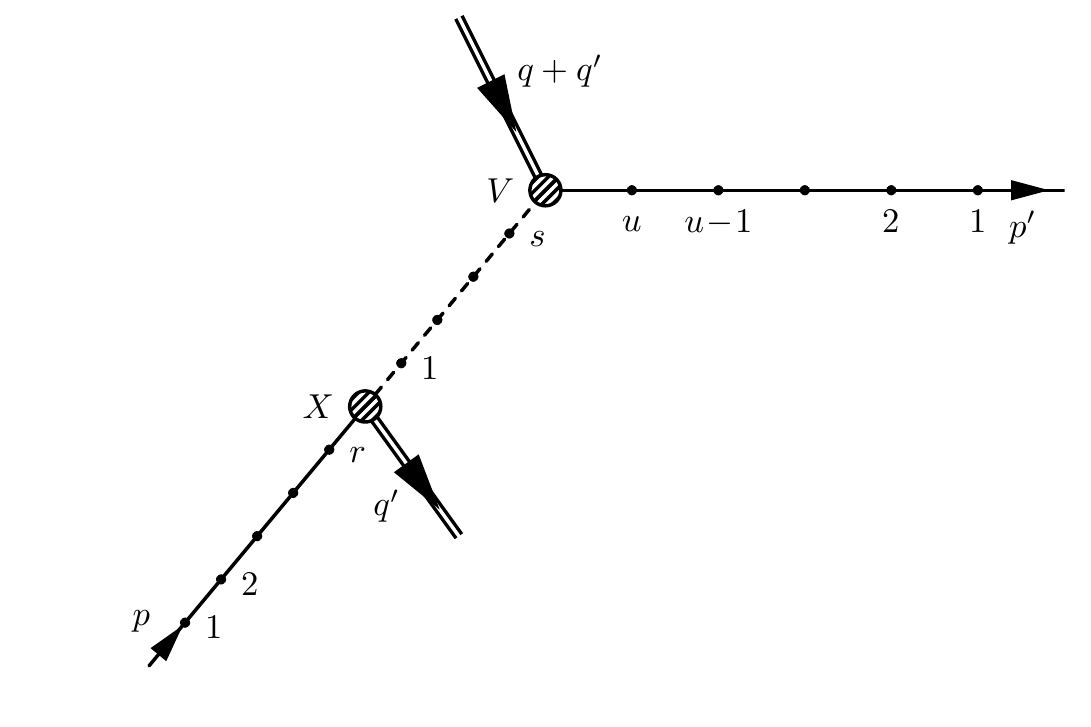}
\caption{\em Defining the $p'$ and $p$ legs for the process, $\chi
\mathcal{F} \to \mathcal{F} \chi$. There are $r$ vertices on the $p$
leg, of which $w$ are on the initial fermion line.} 
\label{fig:darkvertex}
\end{center}
\end{figure}

Our choice is to define the ``$p'$-leg'' to be the final state fermion
line, with the hard momentum $q$ entering at the vertex $V$ of the
initial $\chi$ with the final state fermion and the intermediate
scalar ($\chi$-$\phi$-${\cal{F}}$ vertex), while the ``$p$-leg''
spans both the initial fermion and the intermediate scalar lines; see
Fig.~\ref{fig:darkvertex}. Denote the initial fermion vertex
(${\cal{F}}$-$\phi$-$\chi$) by $X$. There are $u$ vertices on the $p'$
leg, $r$ vertices on the initial fermion line of the $p$ leg and $s$
vertices on the scalar line of the $p$ leg, with $u+s+r=n$.

Hence the momentum of the particle to the {\em right} of the $q^{\rm th}$
vertex on the fermion $p$ leg is $(p + \Sigma_{i=1}^q t_i)$ which
we denote as $p + \Sigma_q$, while the momentum corresponding
to the particle line to the {\em left} of the $q^{\rm th}$ vertex on the
$p'$ leg is $(p' + \Sigma_{i=1}^q l_i)$, which we denote as $p' +
\Sigma_q$. The momentum $q'$ flows out at the bino-fermion-scalar
vertex $X$; hence the momentum of the scalar line just to the right
of vertex $X$ is given by $P \equiv p-q'+\Sigma_r$; hence the
momentum flowing to the {\em right} of the $q^{\rm th}$ vertex on the
scalar line can be expressed as $P + \Sigma_q$. Note that the
momenta are defined so that the sum $p+q = p'$, as in the case discussed
by GY.

With this definition of the vertex $V$ and the $p'$ and $p$ legs, we are
ready to apply the technique of GY to the case of thermal field theories
of binos interacting with charged fermions and scalars.

As in GY, we consider the $n^{\rm th}$ order graph of
Fig.~\ref{fig:darkvertex} and begin by considering the insertion of
an $(n+1)^{\rm th}$ virtual photon to it. Note that the form of the
photon propagator in the thermal case continues to be proportional
to $g_{\mu\nu}$ in the thermal case as well, although otherwise more
complex, and of $2\times 2$ form, as seen from Eq.~\ref{eq:thermalD}
and Eq.~\ref{eq:gammaprop}, thus enabling the separation into $K$ and $G$
photon contributions \`a la GY :
\begin{align} \nonumber
g_{\mu\nu} & = 
	\left\{\strut \left[\strut g_{\mu\nu} - b_k (p_f, p_i)\,k_\mu
	k_\nu \right] + \left[\strut b_k (p_f, p_i) k_\mu k_\nu \right]
	\right\}~, \nonumber \\
 & \equiv \left\{\left[\strut  G\right] +
 \left[K\right] \right\}~,
\label{eq:gmnGK}
\end{align}
with
\begin{align}
b_k(p_f,p_i) = \frac{1}{2} \left[ \frac{(2p_f-k) \cdot
(2p_i-k)}{((p_f-k)^2-m^2)((p_i-k)^2-m^2)} + (k \leftrightarrow -k)
\right]~.
\label{eq:bk}
\end{align}
There are many ways in which the vertices $\mu$ and $\nu$ of the
$(n+1)^{\rm th}$ photon can be inserted into the $n^{\rm th}$ order
graph. This includes
\begin{enumerate}
\item the two vertices of the virtual photon being on {\em different} fermion lines,
\item both vertices of the virtual photon on either of the fermion lines, 
\item both vertices on the intermediate scalar line, and 
\item one of the vertices on either of the fermion lines and the other
on the scalar line.
\end{enumerate}

We consider each of these possibilities, starting with the insertion of
a virtual $K$ photon in the next subsection, and the insertion of a $G$
photon in the subsequent one. The new results in this paper are
essentially from the consideration of the $K$ photon insertions. The
proof that the $G$ photon insertions are finite follows from the
detailed discussions of the corresponding thermal scalar and fermionic
theories in Refs.~\cite{paper1,Indu} and are reproduced here in brief.
Additionally, the case of insertion of real photons is also
strightfoward and the old results apply.  

\subsection{Insertion of virtual $K$ photons}
\label{ssec:kgamma}

We consider an $n^{\rm th}$ order graph as depicted in
Fig.~\ref{fig:darkvertex} and its higher order correction through a
virtual $K$ photon insertion with momentum $k$ leaving/entering the graph
at vertices $\mu$/$\nu$ on the ($p_f$)/($p_i$) legs respectively. Here
$p_f$ and $p_i$ can be one of $p'$ and $p$. (We consider the real photon
contributions later). We express the $n^{\rm th}$ order matrix element
generically as,
\begin{equation}
{\cal{M}}_{n}^{p_f,p_i} = 
{\cal{C}}_{u}^{{\rm fermion}~p'}  \times
{\cal{C}}_{s}^{\rm scalar}  \times
{\cal{C}}_{r}^{{\rm fermion}~p}~,
\end{equation}
where the subscripts on the right hand side simply denote the number
of photon legs attached to the subdiagram. The first term includes the
contribution of the fermion line from the final state fermion to the
initial state bino, including vertex $V$; the second term contains the
contributions of the scalar propagator between vertices $V$ and $X$,
and the third term contains the contribution of the fermion line from
the final bino to the initial fermion, including vertex $X$.

There are three possible types of insertion of the two vertices of the
$(n+1)^{\rm th}$ $K$ photon, depending on the location where the vertices
are inserted as listed above.
\begin{itemize}
\item While the insertion of both $\mu$ and $\nu$ vertices on the $p'$
leg is analogous to the case studied by Grammer and Yennie, the other
two cases differ.

\item This is because the insertion with one vertex on the $p'$ and one
on the $p$ leg includes two possibilities, one where the $\nu$ vertex on
the $p$ leg is on the scalar line and the other where it is on the
initial fermion line (with the $\mu$ vertex being on the $p'$ line in
both cases).

\item Similarly, the case where both $\mu$ and $\nu$ vertices are inserted
on the $p$ leg includes three possibilities, viz., both vertices on the
scalar line, both vertices on the fermion line, and one on each. Where
both vertices are inserted on the same line, care must be taken to avoid
double-counting by ordering the vertices.

\end{itemize}
We see that the case of thermal dark matter is a combination of thermal
fermionic and scalar contributions. The effect of inserting a virtual $K$
photon in all possible ways on a fermion and scalar line were considered
in detail in Refs.~\cite{Indu} and \cite{paper1} respectively; we will
use these results to compute the contributions for the dark matter
interaction at hand. For the sake of completeness, we give the main highlights
of these results in the next section.

\subsubsection{Insertion of virtual $K$ photon vertices into a fermion
or scalar line}
\label{sss:overview}

In both cases, the process considered was the $2 \to 1$ process $\varphi(p)
\gamma^*(q) \to \varphi(p')$ where $\varphi$ is appropriately a charged
fermion or a charged scalar. Consider an $n^{\rm th}$ order graph
with some number of real and virtual photon corrections, and consider
the effect of adding an $(n+1)^{\rm th}$ virtual $K$ photon in all
possible ways. The starting point is to consider the effect of inserting
a single $\mu$ vertex of the virtual $K$ photon adjacent to an arbitrary
vertex $q$ on both a fermion and scalar line and then add all possible
permutations. Details can be found in Refs.~\cite{paper1,Indu}. The main
points to note are the following:

\begin{enumerate}
\item The insertion of a single vertex $\mu$ of a virtual $K$ photon on
a fermion line gives rise to a difference of two terms in the thermal
case, just as in GY at $T=0$. We have (see Eq.~\ref{eq:fid} in Appendix
\ref{app:fidentities}), for an insertion of vertex $\mu$ between
vertices $q$ and $q+1$, 
\begin{align} \nonumber
{\cal{M}}_{n+1} \propto & ~~\bar{u}_{p'} \gamma_{\mu_1} \cdots 
\gamma_{\mu_{q-1}} S\strut^{q\mu}_{p'+\sum_q} \, \slashed{k} \,
	S\strut^{\mu,q+1}_{p'+\sum_q+k} \cdots \Gamma_V
	u_{q+q'}~, \\
	& \propto \bar{u}_{p'} \gamma_{\mu_1} \cdots 
	\gamma_{\mu_{q-1}} \left[S\strut^{q,q+1}_{p'+\sum_q}
	\delta_{t_\mu,t_q+1} - S\strut^{q,q+1}_{p'+\sum_q+k}
	\delta_{t_\mu,t_q} \right]~ \cdots \Gamma_V u_{q+q'}~.
\label{eq:qf}
\end{align}
Here, the propagators $S$ correspond to fermions. The contributions
when the vertex is inserted in all possible ways on a fermion line
cancel term by term in pairs, leaving behind two terms, one which
contains a factor $p'^2-m^2$ which vanishes, and the other which is
proportional to the lower order matrix element (see Ref.~\cite{Indu} for
details). Symbolically, the contributions can be expressed as a sum of
$s+1$ terms with $\mu$ inserted to the left of the vertex labelled $s$,
to the left of vertex $s-1$, etc., all the way to the graph with
$\mu$ inserted to the right of the vertex $1$. We have,
\begin{align} \nonumber
{\cal{M}}_{n+1}^{p',\mu} \sim & 
 = \left\{\strut (p'^2 -m_f^2) \times M_0 + M_1 \right\} \nonumber \\
 &  + \left\{\strut M_2 - M_1 \right\}
 + \left\{\strut \cdots \right\}
 + \left\{\strut M_s - M_{s-1} \right\}
 + \left\{\strut M_{s+1} - M_{s} \right\}~.
\label{eq:p'f}
\end{align}
This demonstrates the term by term cancellation. The left over term is
\begin{align}
M_{s+1} = & \bar{u}_{p'} \gamma_1 S\strut^{t_1,t_2}_{p'+\sum_1} \cdots
S\strut^{t_s,t_V}_{p'+\sum_s} \,\delta_{t_\mu,t_V} \cdots~,
\end{align}
and is proportional to the lower order matrix element, along with the
thermal factor $\delta_{t_\mu,t_V}$. Since the hard vertex is an
external one, $t_V=1$.

\item The insertion of a single vertex $\mu$ on a charged scalar line
has two possibilities: one that the vertex is a new insertion on that
line, leading to the formation of a new 3-point vertex. In this case too
(see Eq.~\ref{eq:sid} of Appendix \ref{app:fidentities}), the result is
the difference of two terms. However, another possibility exists, viz.,
that the photon is inserted at an already existing vertex, converting the
existing 3-point vertex to a 4-point vertex, an option that does not exist
for fermionic QED. A simplification occurs when the contributions from the
insertion of the $\mu$ vertex {\em to the right of the vertex} $q$ and
that from the insertion of the $\mu$ vertex {\em at the vertex} $q$ are
added together to form a so-called ``circled vertex'' \cite{paper1},
denoted ${}_q\mu$. The resulting contribution is again a difference of
two terms, although not as simple and clean a factorisation:
\begin{align}
{\cal M}_{n+1}^{q\mu,{\rm tot}} = & ~ S^{t_V,t_s}_{P+\sum_s}
	\cdots \left[ S\strut^{t_{q-1},t_q}_{P+\sum_{q-1}}
	\delta_{t_\mu,t_q}
	\left(2P+2\Sigma_{q-1}+l_q\right)_{\mu_q} \right. \nonumber \\
 & \left. - S\strut^{t_{q-1},t_q}_{P+\sum_{q-1}+k}
 	\delta_{t_\mu,t_{q-1}}
	\left(2P+2\Sigma_{q-1}+2k+l_q\right)_{\mu_q} \right]
	S\strut^{t_q,t_{q+1}}_{P+\sum_q+k} \cdots 
	S\strut^{t_1,t_X}_{P+\sum_1+k}~, 
\label{eq:circledq}
\end{align}
where $P \equiv p-q'+\Sigma_r$ is the momentum to the right of vertex $X$.
Here each $S$ corresponds to the scalar propagator with the appropriate
thermal indices and momentum. This is the result for scalar thermal QED
corresponding to Eq.~\ref{eq:qf} above for the fermionic case. Again,
inserting the vertex $\mu$ in all possible ways on the scalar line gives
contributions that cancel term by term giving rise to a great deal of
simplification and a result that is similar to Eq.~\ref{eq:p'f}. An
examination of the thermal factors shows that the left-over term
arises from the insertion of the virtual $K$ photon vertex adjacent to
the hard vertex $V$.

Note that, in contrast to the initial and final fermion lines which
have an outermost component that is on-shell (with momenta $p$ and $p'$
respectively), neither end of the scalar line is on-shell. Hence, the
earlier results on thermal scalar QED \cite{paper1} need to be modified
to bring back the terms proportional to $P^2-m^2$ that vanished in the
earlier calculation. In the calculation that follows we therefore list
separately the ``on-shell'' contribution that was already calculated in
the earlier papers, as well as the new ``off-shell'' contribution due
to these changes.

\item In the case when both $\mu$ and $\nu$ vertices are on the {\em
same} leg, they are ordered with $\nu$ always to the left of $\mu$, to
avoid double counting. In addition, for scalars, it is possible that two
photons are inserted at the {\em same} vertex, giving rise to 4-point
seagull vertices, in contrast to the fermionic case where only 3-point
vertices exist. When both $\mu$ and $\nu$ vertices are inserted at the
{\em same} point, we get 4-point tadpole diagrams, which again do not
exist for fermions.  This is the most complex case of all. We do not
highlight details of this calculation but simply refer the reader to
the original papers \cite{paper1,Indu}. We note only that the cancellation
occurs again, leaving behind a term that is proportional to the lower
order matrix element, but in this case the corresponding thermal factors
of the left-over term are $\delta_{t_\mu,t_1} \delta_{t_\nu,t_1}$.

\item We summarise and highlight the dependence on the thermal factors
as follows. When the vertices of the $(n+1)^{\rm th}$ $K$ photon are
on different legs, the thermal factor of the left-over term that is
proportional to the lower order matrix element depends on the thermal type
of the hard vertex $V$, viz., $\delta_{t_\mu,t_V} \delta_{t_\nu,t_V}$.
When the vertices of the $(n+1)^{\rm th}$ $K$ photon are on the same ($p'$
or $p$) leg, the thermal factor of the left-over term that is proportional
to the lower order matrix element depends on the thermal type of the
outermost line of that leg (with momentum $p'$ or $p$ respectively),
viz., $\delta_{t_\mu,t_1} \delta_{t_\nu,t_1}$. The hard vertex $V$
involves participation of external/observable particles; the outermost
lines of $p$ and $p'$ legs are external lines as well; hence both are
of thermal type 1. Thus the term that survives picks out only the (11)
component of the inserted photon propagator, $i {\cal D}^{ab}_{\mu\nu}
(k)$. This is critical to achieve cancellation of IR divergent terms
between the virtual and real photon contributions.

\end{enumerate}

We will now apply these results to examine the contributions to the
graph in Fig.~\ref{fig:darkvertex}. We begin with the case when the two
vertices are on the $p'$ leg.

\subsubsection{Insertion of both vertices of the virtual $K$ photon on the
$p'$ leg}

This is the most straightforward part of the calculation and simply
follows from the results for insertion of the $(n+1)^{\rm th}$
$K$-photon on the $p'$ leg as given in Ref.~\cite{Indu} which deals with
thermal fermionic QED. Both vertices are inserted in all possible ways
in the lower order graph, keeping $\nu$ always to the left of $\mu$ in
order to avoid double counting. The resulting contribution from all
possible symmetric contributions is a term proportional to the lower
order matrix element. Since the insertion only affects the $p'$ leg, the
parts of the matrix element involving the scalar and initial fermion $p$
legs are unchanged and we get,
\begin{equation}
{\cal M}\strut^{p'p', K\gamma}_{n+1} = +ie^2
	\int \frac{{\rm d}^4 k}{(2\pi)^4} \, \delta_{t_\mu,t_1} \,
	\delta_{t_\nu,t_1} \, b_k(p',p') \, D^{t_\mu,t_\nu} (k)~{\cal M}_n~.
\label{eq:Kgp'p'}
\end{equation}
Since $t_1=1$ necessarily, it depends only on the $D^{11}$ photon
propagator, as before.

\subsubsection{Insertion of virtual $K$ photon vertices separately on $p'$
and $p$ legs}

We now consider the case when the $\mu$ vertex is inserted on the $p'$
leg and the $\nu$ vertex is inserted on the $p$ leg. The matrix
element can be represented as,
\begin{equation}
{\cal{M}}_{n+1}^{p',p} \sim
{\cal{C}}_{u+1}^{{\rm fermion}~p',\mu}  \times 
{\cal{C}}_{r+s+1}^{p,\nu}~.
\label{eq:pp'}
\end{equation}
Again, the subscript simply indicates the number of photon legs on the
corresponding subdiagram, while the superscripts indicate the location
where the vertices $\mu$ and $\nu$ are inserted.

\paragraph{Insertion of the $\mu$ vertex on the $p'$ leg}

The contributions from inserting $\mu$ on the $p'$ leg and $\nu$ on the
$p$ leg factorise and can be separately considered. The contribution
from inserting a single vertex $\mu$ from a $K$ photon insertion on the
fermion $p'$ leg has been considered in detail in Ref.~\cite{Indu} and is
proportional to the insertion of a factor $\slash{\!\!\!k}$ at the $\mu$
vertex; see Eq.~\ref{eq:qf} above. On including all possible insertions,
the contributions cancel term by term in pairs, leaving behind only one
term. The relevant part of the matrix element involving only the $p'$
leg and the vertex $\mu$, inserted on it in all possible ways is given by,
\begin{equation}
{\cal{C}}_{u+1}^{{\rm fermion}~p',\mu}  \sim \bar{u}_{p'} \gamma_{\mu_1}
S\strut^{t_1 t_2}_{p'+\sum_1} \cdots (\hbox{no }k) \cdots 
S\strut^{t_s t_V}_{p'+\sum_s} \, \delta_{t_\mu,t_V} \, \Gamma_V
\, u_{q+q'}~,
\label{eq:fermionp'}
\end{equation}
where ``no $k$" implies that all the corresponding terms have no $k$
dependence in them; $\Gamma_V$ represents the fermion-bino-scalar
vertex. Here we have suppressed various overall factors for clarity and
will put them back later. Hence we see that apart from the thermal factor
$\delta_{t_\mu,t_V}$, this part of the ${\cal{M}}_{n+1}$ matrix element
is proportional to the corresponding part of the lower order matrix
element, ${\cal{M}}_{n}$.

\paragraph{Insertion of the $\nu$ vertex on the $p$ leg}
\label{subsub:scalar_notn}

There are two contributions:
\begin{enumerate}
\item Insertion of the vertex $\nu$ on the scalar line,
\item Insertion of the vertex $\nu$ on the inital fermion line.
\end{enumerate}
We denote their contributions to the resulting matrix element as,
\begin{equation}
{\cal{C}}_{r+s+1}^{p,\nu} \sim
\left[
{\cal{C}}_{s+1}^{{\rm scalar},\nu}  \times
{\cal{C}}_{r}^{{\rm fermion}~p}(q_i) +
{\cal{C}}_{s}^{\rm scalar}(q_i \to q_i +k)  \times
{\cal{C}}_{r+1}^{{\rm fermion}~p,\nu} \right]~.
\label{eq:pp'2}
\end{equation}
When the $\nu$ vertex is inserted on the initial fermion line, the
momentum $k$ flows through the entire scalar line from the initial to the
final fermion lines, but otherwise there are no new insertions on the
scalar line. Hence the only effect of inserting the additional virtual
photon on the fermion line is that every momentum on the scalar line is
shifted by an additional factor, $k$, so that this portion of the matrix
element is the same as the relevant lower order matrix element with
shifted momenta:
\begin{equation}
{\cal{C}}_{s}^{\rm scalar} =
{\cal{C}}_{s}^{\rm scalar}(q_i \to q_i +k)~.
\label{eq:scalarshift}
\end{equation}
When the $\nu$ vertex is inserted on the scalar line, the initial
fermion line remains unaffected, since the additional momentum $k$ does
not flow through it. Hence every intermediate propagator, $q_i$, remains
the same as in the lower order graph, and the relevant portion of
this matrix element is in fact the same as that of the lower order graph:
\begin{equation}
{\cal{C}}_{r}^{{\rm fermion}~p} = 
{\cal{C}}_{r}^{{\rm fermion}~p} (q_i \to q_i)~.
\label{eq:fermionnoshift}
\end{equation}
The remaining terms in Eq.~\ref{eq:pp'2} are calculated below.

\paragraph{Insertion of the $\nu$ vertex of virtual $K$ photon on the
scalar line}

The contribution from inserting the $\nu$ vertex from a virtual $K$
photon insertion on the scalar part of the $p$ leg has been discussed in
detail in Ref.~\cite{paper1}. For convenience, we relabel the momenta of
the lines between the vertices $V$ and $X$ as $P+\Sigma_s$, $P+\Sigma_{s-1}$,
$\cdots$, $P+\Sigma_1$ respectively, with $P \equiv p-q'+\Sigma_r$ being
the momentum of the propagator to the right of the vertex $X$, where
$k_1, k_2, \ldots, k_r$ are the photon momenta entering the $r$
vertices on the $p$ fermion leg.

The major difference between the earlier calculation and this one is
that here the scalar line is completely off-shell, unlike earlier,
where it was either the final or initial leg, with the outermost line
on-shell. Hence we have to account for this difference in the current
calculation. Apart from the inclusion of the scalar propagator between
the vertices $1$ and $X$ since this line is no longer on-shell, we must
take into account the fact that $P$ itself is off-shell. Hence, a term
in the earlier calculation for scalar QED which vanished because $P^2 =
m^2$ is now not zero. We get,
\begin{align}
\nonumber {\cal{C}}_{s+1}^{{\rm scalar},\nu} & \sim S\strut^{t_X,t_1}_{P}
               (2P+l_1)_{\mu_1} S\strut^{t_1,t_2}_{P+\sum_1}
               \cdots
	S\strut^{t_s,t_V}_{P+\sum_s} \left[\delta_{t_\nu,t_V}
	\right] \nonumber \\
 & ~~~~~ -S\strut^{t_X,t_1}_{P+k} (2P+2k+l_1)_{\mu_1} \cdots
	S\strut^{t_s,t_V}_{P+k+\sum_s} \left[\delta_{t_\nu,t_X}
	\right]~.
\label{eq:MnKp'T}
\end{align}
The first term has no $k$ dependence and is therefore proportional to
the corresponding part of the lower order matrix element; note that this
contribution would have vanished if $P$ was on-shell. We see that there
is an extra term, with $k$ dependence in each part of the contribution,
which arises when the $\nu$ vertex is just to the right of vertex
$X$. In contrast to the corresponding result in Ref.~\cite{paper1}
where this contribution was independent of $k$ and proportional to
the lower order matrix element, here every momentum is shifted by $k$
since this momentum runs through practically the entire scalar leg,
from the vertex $V$ to the vertex $\nu$ just to the left of $X$. We will
see below that this extra term cancels against the contribution when the
$\nu$ vertex is inserted on the fermion $p$ line.

\paragraph{Insertion of the $\nu$ vertex of the virtual $K$ photon on
the fermion $p$ line}

The contribution from inserting a single vertex $\nu$ from a virtual $K$
photon insertion on the fermionic part of the $p$ leg has been
calculated in Ref.~\cite{Indu}; since the outermost leg is on-shell,
$p^2 = m^2$, we can apply the earlier results with just the replacement of
the appropriate vertex factor at the vertex $X$. As before, terms from
all possible insertions of this vertex cancel in pairs, and we have,
\begin{equation}
{\cal{C}}_{r+1}^{{\rm fermion}~p,\nu} \sim \bar{u}_{q'} \Gamma_X
	S\strut^{t_X t_r}_{p+\sum_r} \delta_{t_\nu,t_X}
	\gamma_r \cdots (\hbox{no }k)~u_p~,
\end{equation}
with the corresponding scalar contribution having the factor $k$ added
to the momentum of each propagator: $q_i \to q_i +k$, as mentioned in
Eq.~\ref{eq:scalarshift} above. Both these insertions contribute to
the matrix element in Eq.~\ref{eq:pp'2}. Substituting,
Eq.~\ref{eq:pp'2} simplifies due to cancellations as shown below to,
\begin{align} \nonumber
{\cal{C}}_{r+s+1}^{p,\nu} & \sim \left[
	\left\{S\strut^{t_X,t_1}_{P} (2P+l_1)_{\mu_1}
	S\strut^{t_1,t_2}_{P+\sum_1} \cdots
        S\strut^{t_s,t_V}_{P+\sum_s} \left[\delta_{t_\nu,t_V} 
	\right] \right. \right. \nonumber \\
  & ~~~~~ \left. -S\strut^{t_X,t_1}_{P+k} (2P+2k + l_1)_{\mu_1} \cdots
         S\strut^{t_s,t_X}_{P+k+\sum_s} \left[\delta_{t_\nu,t_X}
 	\right]~ \right\} \times
 	\left\{ \bar{u}_{q'} \Gamma_X
 	S\strut^{t_X,t_r}_{p+\sum_r} \gamma_r \cdots
 	(\hbox{no }k)~u_p \right\} \nonumber \\
  & ~~~~~ + \left. \left\{S\strut^{t_X,t_1}_{P+k} (2P+2k + l_1)_{\mu_1} \cdots
         S\strut^{t_s,t_V}_{P+k+\sum_s} \right\} \times 
 	\left\{ \bar{u}_{q'} \Gamma_X
 	S\strut^{t_X,t_r}_{p+\sum_r} \delta_{t_\nu,t_X} \gamma_r \cdots
 	(\hbox{no }k)~u_p \right\} \right]~, \nonumber \\
 & \sim  \left\{S\strut^{t_X,t_1}_{P} (2P+l_1)_{\mu_1}
	S\strut^{t_1,t_2}_{P+\sum_1} \cdots
        S\strut^{t_s,t_V}_{P+\sum_s} \left[\delta_{t_\nu,t_V}
	\right] \right\} \times
 	\left\{ \bar{u}_{q'} \Gamma_X
 	S\strut^{t_X,t_r}_{p+\sum_r} \gamma_r
	\cdots (\hbox{no }k)~u_p \right\}~.
\label{eq:pnu}
\end{align}

\subsubsection{Matrix element for insertion of virtual $K$ photon on
different legs}

Substituting for Eqs.~\ref{eq:pnu} and \ref{eq:fermionp'},
${\cal{M}}_{n+1}^{p',p}$ in Eq.~\ref{eq:pp'} evaluates to,
\begin{align}
{\cal{M}}_{n+1}^{p',p} \sim &
\left\{\bar{u}_{p'} \gamma_{\mu_1}
S\strut^{t_1 t_2}_{p'+\sum_1} \cdots (\hbox{no }k) \cdots 
S\strut^{t_s t_V}_{p'+\sum_s} \, \delta_{t_\mu,t_V} \, \Gamma_V
\, u_{q+q'} \right\} \times \nonumber \\
 & \left\{S\strut^{t_X,t_1}_{P} (2P+l_1)_{\mu_1}
	S\strut^{t_1,t_2}_{P+\sum_1} \cdots
        S\strut^{t_s,t_V}_{P+\sum_s} \left[\delta_{t_\nu,t_V}
	\right] \right\} \times
 	\left\{ \bar{u}_{q'} \Gamma_X
 	S\strut^{t_X,t_r}_{p+\sum_r} \gamma_r
	\cdots (\hbox{no }k)~u_p \right\}~, \nonumber \\
 & \propto {\cal{M}}_n \times \left[\delta_{t_\mu,t_V}\right] \,
 \left[\delta_{t_\nu,t_V}\right]~;
\label{eq:pp'ans}
\end{align}
hence the total contribution is proportional to the lower order matrix
element. Note that the propagators may correspond to either fermion or
scalar ones, as appropriate. Putting back the overall factors, including
parts of the photon propagator that had been omitted, we have,
\begin{align}
{\cal M}\strut^{p'p,K\gamma}_{n+1} & = -ie^2 \int \frac{{\rm d}^4
k}{(2\pi)^4} \, \delta_{t_\mu,t_V} \, \delta_{t_\nu,t_X} b_k(p',p)
D^{t_\mu,t_\nu} (k) \times {\cal M}_n~.
\label{eq:Kgp'p}
\end{align}
Note that for thermal QED we had obtained \cite{Indu,paper1} thermal
terms that were proportional to $\delta_{t_\mu,t_V} \delta_{t_\nu,t_V}
D^{t_\mu,t_\nu} (k)$, and hence to the (11) element of the thermal photon
propagator, $D^{11}(k)$, since there was a single hard vertex, $V$. We
obtain a similar result here as well, since the bino particles at both the
$V$ and $X$ vertices are external particles and hence of thermal type-1;
that is, the matrix element is again proportional to the (11) element of
the photon propagator, $D^{11}(k)$, as before. This result also holds
for arbitrary number of intermediate scalar and fermion propagators on
the $p$ leg (with arbitrary corresponding emissions of bino particles
at each of these vertices) since the bino is always on-shell.

\paragraph{Discussion on the nature of the cancellation}

In the case
where both the vertices of the $(n+1)^{\rm th}$ photon are inserted
on the $p'$ leg, the term-by-term cancellation, leaving behind just
one term proportional to the lower order matrix element, as shown in
Eq.~\ref{eq:Kgp'p'}, occurs just as in the case of thermal fermionic
QED discussed in Ref.~\cite{Indu}. However, when the two vertices are
inserted on different legs, the result shown in Eqs.~\ref{eq:pnu}
and \ref{eq:pp'ans} involves a double cancellation. This can be
understood as follows. The $\mu$ vertex insertion on the $p'$ leg is
straightforward. However, there are two parts to the $p$ leg, viz., the
scalar propagator, and the initial fermion $p$ line. A set of term-by-term
cancellations occurs when the $\nu$ vertex is inserted in all possible
ways on the fermion $p$ line. However, even though this part of the matrix
element is then proportional to the lower order matrix element, the part
of the matrix element from the scalar propagator that multiplies this,
contains the $k$ dependence since this extra momentum flows through it. On
the other hand, the contribution when the $\nu$ vertex is inserted in all
possible ways on the scalar propagator has a term by term cancellation,
leaving behind {\em two} terms. Due to the presence of the thermal
indices, it is easy to see from Eq.~\ref{eq:MnKp'T} that one term arises
from the $\nu$ insertion adjacent to vertex $V$ and the other one from an
insertion adjacent to vertex $X$. We know from the discussion in Section
\ref{sss:overview} that insertions on two different legs (in the purely
fermionic or purely scalar case) results in a term that is proportional
to $\delta_{t_\nu,t_V}$. The fact that there is an additional term with
$\delta_{t_\nu,t_X}$ indicates the incomplete cancellation owing to the
relevant momentum (to the right of vertex $X$) not being on-shell. All
possible insertions of $\nu$ on the remaining part of the $p$ leg, viz.,
on the fermionic $p$ line, allow a term by term cancellation that leaves
just one term. As per the discussion in Section \ref{sss:overview},
this should arise from insertion adjacent to the hard vertex which is
$X$ in this case. Hence the corresponding term is proportional to the
thermal factor $\delta_{t_\nu,t_X}$ and this term cancels the left-over
term from all possible insertions on the scalar part of the $p$ leg.

It can also be seen that nowhere was the exact form of the bino vertex
required, either at vertex $V$ or vertex $X$. The term-by-term
cancellation occurs independently of the exact nature of this
interaction. Hence the analysis is ``blind" to the {\em precise
structure} of the hard process at vertices $V$ and $X$.

\subsubsection{Insertion of both vertices of the virtual $K$ photon on the
$p$ leg}

Since the $p$ leg comprises both the scalar line as well as the initial
fermion line, this is the most complex part of the calculation and has
three components with three different types of insertion of the vertices
of the virtual $K$ photon:
\begin{enumerate}
\item Both vertices on the fermion $p$ line,
\item One vertex on the scalar line and one on the fermion $p$ line,
\item Both vertices on the intermediate scalar line.
\end{enumerate}
Since the contribution from the $p'$ leg is unaffected by the insertion,
we denote these contributions as
\begin{align} \nonumber
{\cal{M}}_{n+1}^{p,p} & \sim
{\cal{C}}_{u}^{{\rm fermion}~p'}  \times \left[
{\cal{C}}_{s}^{\rm scalar}  \times
{\cal{C}}_{r+2}^{{\rm fermion}~p;\mu,\nu} + 
{\cal{C}}_{s+1}^{\rm scalar;\mu}  \times
{\cal{C}}_{r+1}^{{\rm fermion}~p;\nu} \right. \nonumber \\
 & ~~~~~~~~~~~ \left. +  {\cal{C}}_{s+2}^{{\rm scalar};\mu, \nu}  \times
{\cal{C}}_{r}^{{\rm fermion}~p} \right]~.
\label{eq:pp}
\end{align}
We will study each in turn. Again, we highlight that neither of the
outermost legs of the scalar line is on shell; hence we have to bring
back the terms which were dropped on account of setting $P^2 -m^2 =
0$. We refer to these extra terms as ``off-shell'' contributions that
occur over and above the earlier results which are labelled as the
``on-shell'' part. We also have to include the propagator corresponding to
the outermost part of the scalar line as an overall multiplicative factor
for all terms since it is not on-shell; this is a trivial modification.

We begin with the insertion of both vertices on the fermion part of the
$p$ leg.

\paragraph{Insertion of both $\mu$, $\nu$ vertices of the $K$ photon
on the fermionic $p$ leg}

This involves inserting both vertices of the virtual $K$ photon on a
fermion line having $r$ vertices, in all possible ways. There is no
$k$ dependence in the rest of the diagram (the final $p'$ leg or the
intermediate scalar line). This has been calculated in Ref.~\cite{Indu}
for fermionic thermal QED and the result is proportional to the
lower order matrix element. However, as per GY, we have to remove
a term corresponding to the insertion of $\nu$ and $\mu$ just to the
right of vertex $1$ on the fermion line, to account for wave function
renormalisation. Hence the net contribution of this insertion is zero,
as was the corresponding case for purely fermionic thermal QED
\cite{Indu}:
\begin{align} \nonumber
{\cal{C}}_{r+2}^{{\rm fermion}~p;\mu,\nu}  & = 0~; \nonumber \\
\hbox{Hence, } {\cal{C}}_{s}^{\rm scalar}  \times
{\cal{C}}_{r+2}^{{\rm fermion}~p;\mu,\nu}  & = 0~.
\label{eq:ppff}
\end{align}

\paragraph{Insertion of the $\mu$ vertex of the $K$ photon on the
scalar part and the $\nu$ vertex on the fermionic part of the $p$ leg}

As in the case when the $\nu$ vertex was on the fermionic $p$ leg
and the $\mu$ vertex is on the scalar leg, the contributions due to the
two vertex insertions factorise and can be independently calculated.

For the scalar line, the $\mu$ vertex is again inserted in all possible
ways on the scalar line. There are $s+1$ contributions, as discussed
in Ref.~\cite{paper1}, with $s$ diagrams each having one circled vertex,
${}_q\mu$, $q = 1, \cdots s$, and one diagram with vertex $\mu$ inserted
to the right of vertex $X$; see Ref~\cite{paper1}. Again we note
that neither end of the scalar line is on-shell.  The contributions
from each insertion can be written as a difference of two terms,
analogous to the fermionic result in Eq.~\ref{eq:p'f},
\begin{eqnarray} \nonumber
{\cal{C}}_{s+1}^{{\rm scalar}; \mu} & = & 
\left\{\left[ S\strut^{t_V,t_1}_{P} \delta_{t_\mu,t_1}
 	(2P+l_1)_{\mu_1} - S\strut^{t_V,t_1}_{P+k} \delta_{t_\mu,t_V}
	(2P+2k + l_1)_{\mu_1} \right] \right.  \times \nonumber \\
 & & 	 \left. ~~~~~~~~ S\strut^{t_1,t_2}_{P+\sum_1+k} 
	(2P+2k + 2\Sigma_1+l_2)_{\mu_2} \cdots \right\}
	 \nonumber \\
 	& & + \left\{ S\strut^{t_V,t_1}_{P} (2P+l_1)_{\mu_1}
	\left[S\strut^{t_1,t_2}_{P+\sum_1}
	\delta_{t_\mu,t_2} (2P+2\Sigma_1+l_2)_{\mu_2} 
	\right. \right. \nonumber \\
 & & ~~~~~~~~ \left. \left. -S\strut^{t_1,t_2}_{P+\sum_1+k}
 \delta_{t_\mu,t_1} (2P+2\Sigma_1+2k+l_2)_{\mu_2} \right]
	\cdots \right\} \nonumber \\ 
& & + \left\{\strut \cdots \right\} \nonumber \\
 	& & + \left\{ S\strut^{t_V,t_1}_{P} (2P+l_1)_{\mu_1}
	S\strut^{t_1,t_2}_{P+\sum_1} \cdots
        \left[S\strut^{t_{s-1},t_s}_{P+\sum_{s-1}}
 	\delta_{t_\mu,t_s} (2P+2\Sigma_{s-1}+l_s)_{\mu_s}
	\right. \right. \nonumber \\
 & & \left. \left. + S\strut^{t_{s-1},t_s}_{P+\sum_{s-1}+k}
 	\delta_{t_\mu,t_{s-1}} (2P+2\Sigma_{s-1}+2k+l_s)_{\mu_s}
         \right] \cdots \right\} \nonumber \\
 & & + \left\{ S\strut^{t_V,t_1}_{P} (2P+l_1)_{\mu_1}
 	S\strut^{t_1,t_2}_{P+\sum_1} \cdots
        \left[S\strut^{t_s,t_X}_{P+\sum_s} \delta_{t_\mu,t_X} 
         - S\strut^{t_s,t_X}_{P+\sum_s+k} \delta_{t_\mu,t_s} 
	\right] \right\}~, \nonumber \\
 & \equiv & \left\{\strut M_1 - M_0 \right\}
 + \left\{\strut M_2 - M_1 \right\}
 + \left\{\strut \cdots \right\}
 + \left\{\strut M_s - M_{s-1} \right\}
 + \left\{\strut M_{s+1} - M_{s} \right\}~.
\label{eq:sT}
\end{eqnarray}
It is seen that the terms cancel in pairs, leaving behind two terms,
$M_0$ and $M_{s+1}$. In contrast, the result when a single vertex $\mu$
was inserted in all possible ways on a scalar leg with the external leg
having a momentum $P$ for purely thermal scalar fields \cite{paper1},
it was found to be proportional to just the $M_{s+1}$ term (without the
initial factor, $S^{t_V,t_1}_{P}$); the $M_0$ term evaluated to zero
since $P$ was on-shell.

This factor is multiplied by the part of the matrix element arising
from the fermion part of the $p$ leg. Applying the generalised Feynman
identities defined in Appendix \ref{app:fidentities}, insertion of the
$\nu$ vertex in all possible ways into the fermionic part of the $p$
leg gives sets of terms that cancel against each other. This has been
calculated in the thermal fermionic QED case in Ref.~\cite{Indu}; we
simply use the result obtained in this paper to get,
\begin{equation}
{\cal{C}}_{r+1}^{{\rm fermion}~p;\nu} \sim
 	\left\{ \bar{u}_{q'} \Gamma_X
 	S\strut^{t_X,t_r}_{p'+\sum_r} \delta_{\nu,X} \gamma_r \cdots
 	(\hbox{no }k)~u_p ~\right\}~.
\end{equation}
This is again proportional to the corresponding part of the lower order
matrix element since it is independent of $k$. Combining both terms,
we obtain the product,
\begin{eqnarray} \nonumber
{\cal{C}}_{s+1}^{\rm scalar;\mu} \times
{\cal{C}}_{r+1}^{{\rm fermion}~p;\nu} & \sim &
	\left[ - S\strut^{t_V,t_1}_{P+k} ~\delta_{t_\mu,t_V}
	(2P+2k + l_1)_{\mu_1} \right.  \times \nonumber \\
  & & 	 \left. ~~~~~~~~ S\strut^{t_1,t_2}_{P+\sum_1+k} 
 	(2P+2k + 2\Sigma_1+l_2)_{\mu_2} \cdots \right.
 	 \nonumber \\
  & & + \left. S\strut^{t_V,t_1}_{P} (2P+l_1)_{\mu_1}
  	S\strut^{t_1,t_2}_{P+\sum_1} \cdots
         S\strut^{t_s,t_X}_{P+\sum_s} ~\delta_{t_\mu,t_X} 
 	\right]~ \nonumber \\
  & & 	\times \left\{ \bar{u}_{q'} \Gamma_X
  	S\strut^{t_X,t_r}_{P+\sum_r} ~\delta_{\nu,X} \gamma_r \cdots
  	(\hbox{no }k) u_p ~\right\}~.
\label{eq:ppsf}
\end{eqnarray} 
The term corresponding to $M_0$ in the earlier calculation was zero
owing to the presence of the term $P^2 - m^2 =0$. Note that $M_{s+1}$
is independent of $k$ so that its contribution is proportional to the
lower order matrix element; however, the extra term $M_0$ is not, and
so the factorisation is not yet obtained.

\paragraph{Insertion of both $\mu$, $\nu$ vertices of the $K$ photon 
on the scalar part of the $p$ leg}

As in the case of pure scalar thermal QED, this is a more involved
calculation since there are both 3-point and 4-point vertices for scalars;
also, apart from seagull-type insertions with 2-scalar-2-photon lines
at a vertex, there can also be tadpole-type diagrams with the two photon
lines at the vertex forming a loop. We will briefly outline and use the
results obtained earlier in Ref.~\cite{paper1} and merely highlight the
extra ``off-shel'' contributions. As discussed in Ref.~\cite{paper1},
there are four sets of contributing diagrams:

\begin{enumerate}
\item Set I, with circled vertices ${}_q\mu$ and ${}_q\nu$ at both
$\mu$ and $\nu$ insertions, where $q$ are any of the already existing
vertices, $q=1, \cdots, s$.

\item Set II, with circled vertices ${}_q\mu$ only at $\mu$, with $\nu$
to the right of the special vertex, $X$.

\item Set III, with all 4-point vertex insertions of $\nu$, so that
$\nu$ is inserted at any of the already existing $q$ vertices, $q=\nu$,
with $\mu$ immediately adjacent to $\nu$.

\item Finally, Set IV, which is a set of ${}_\nu\mu$ circled vertices
that includes all tadpole insertions, $\mu=\nu$, as well.

\end{enumerate}

Note that the calculation in Ref.~\cite{paper1} was for insertions on the
scalar $P$ leg where the external leg, $P$ was on-shell. Here this is
not the case here and this gives rise to extra contributions. It turns
out that it is possible to express the result of these contributions
as the sum of two parts: one, which is the same as the result in
Ref.~\cite{paper1} (with the usual additional propagator prefactor
$S^{t_V,t_1}_{p'}$), referred to as the ``on-shell'' contribution,
and the second, which are the terms arising from the fact that $P$ is
not on-shell which we label the ``off-shell'' contribution. Note that
this is simply a convenient assignment; all terms of the part labelled
``on-shell'' are not explicitly dependent on $P$ being on-shell;
rather they are blind to the off-shell or on-shell nature of $P$. It is
the ``off-shell'' part that arises strictly owing to the off-shell
nature of $P$. We now explicitly list the extra contributions to the
various sets below, and refer the reader to Ref.~\cite{paper1} for the
``on-shell'' contribution.

The contribution from Set I is given by,
\begin{eqnarray} \nonumber
{\cal{C}}_{s+2}^{{\rm scalar}; \mu,\nu, I} & = &  
	S\strut^{t_V,t_1}_{P} \left\{{\hbox{On-shell}} \right\} \nonumber \\
 & &  + S\strut^{t_V,t_1}_{P+k} ~\delta_{t_\mu,t_V} (2P+2k +l_1)_{\mu_1}  
 	\left\{ -S\strut^{t_1,t_2}_{P+\sum_1}
 ~\delta_{t_1,t_\nu} (2P+2\Sigma_1+l_2)_{\mu_2} \cdots (\hbox{no } k)
\right. \nonumber \\
 & & + \left. S\strut^{t_1,t_2}_{P+\sum_1+k}
 	(2P+2\Sigma_1+2k+l_2)_{\mu_2} 
	\cdots (\hbox{all } k) ~\delta_{t_\nu,t_s} \right\}~, 
\label{eq:I}
\end{eqnarray} 
where ``no $k$" and ``all $k$" refer to the fact that none or all of the
vertex and propagator terms in the sequence are $k$-dependent and the
``off-shell'' contributions only have been explicitly listed.
Similarly, the contribution from Set II can be expressed as,
\begin{eqnarray} \nonumber
{\cal{C}}_{s+2}^{{\rm scalar}; \mu,\nu, II} & = &  
	S\strut^{t_V,t_1}_{P} \left\{{\hbox{On-shell}} \right\} \nonumber \\
 & &  - S\strut^{t_V,t_1}_{P+k} ~\delta_{t_\mu,t_V} (2P+2k +l_1)_{\mu_1}  
 	S\strut^{t_1,t_2}_{P+\sum_1+k}
 	(2P+2\Sigma_1+2k+l_2)_{\mu_2} 
	\cdots (\hbox{all } k) \times \nonumber \\
 & & \left[
 	S\strut^{t_s,t_X}_{P+\sum_s} ~\delta_{t_\nu,t_s}
 	- S\strut^{t_s,t_X}_{P+k+\sum_s} ~\delta_{t_\nu,t_X}
	\right]~.
\label{eq:II}
\end{eqnarray} 
The contribution from Set III can be expressed as,
\begin{eqnarray} \nonumber
{\cal{C}}_{s+2}^{{\rm scalar}; \mu,\nu, III} & = &  
	S\strut^{t_V,t_1}_{P} \left\{{\hbox{On-shell}} \right\} \nonumber \\
 & &  - S\strut^{t_V,t_1}_{P+k} ~\delta_{t_\mu,t_V} (-2k)_{\mu_1}  
 	~\delta_{t_\nu,t_1} S\strut^{t_1,t_2}_{P+\sum_1}
	\cdots (\hbox{no } k) ~
 	S\strut^{t_s,t_X}_{P+\sum_s}~.
\label{eq:III}
\end{eqnarray} 
As in the case of pure scalar thermal QED \cite{paper1}, the tadpole
contributions in Set IV cancel the $k^2$ dependent terms in the remaining
part of Set IV. This is again crucial to the factorisation since $k^2$
terms are IR finite (both at zero and finite temperature) but will spoil
the factorisation and subsequent exponentiation at all orders of the IR
divergent coefficients. We have,
\begin{eqnarray} \nonumber
{\cal{C}}_{s+2}^{{\rm scalar}; \mu,\nu, IV} & = & 
	S\strut^{t_V,t_1}_{P} \left\{{\hbox{On-shell}} \right\} \nonumber \\
 & &   +~ 0~.
\end{eqnarray} 
Note that the calculation in Ref.~\cite{paper1} did not include the
self-energy correction diagrams on the outermost part of the $p$ scalar
leg, to account for wave function renormalisation. Here, however, the
scalar $P$ leg is off-shell. Adding back these terms to Set IV, we get,
\begin{eqnarray} \nonumber
{\cal{C}}_{s+2}^{{\rm scalar}; \mu,\nu, IV} & = & 
	S\strut^{t_V,t_1}_{P} \left\{{\hbox{On-shell}} \right\}
 +~ {\cal{M}}_{n+1}^{{\rm scalar}; \mu,\nu, {\rm Self}} \nonumber \\
 & = & S\strut^{t_V,t_1}_{P} \left\{{\hbox{On-shell}} \right\} \nonumber \\
 & & + \left\{
 	S\strut^{t_V,t_\nu}_{P} ~\delta_{t_\mu,t_\nu} (2P\cdot k)\,
 	S\strut^{t_\nu,t_1}_{P} - ~\delta_{t_\mu,t_V}
	\left[S\strut^{t_V,t_1}_{P} ~\delta_{t_\nu,t_V}
	- S\strut^{t_V,t_1}_{P+k} ~\delta_{t_\nu,t_1} \right] \right\}
	  \times \nonumber \\
 & & ~~~~~~~~~~ (2P+l_1)_{\mu_1} \, S\strut^{t_1,t_2}_{P+\sum_1}
 	\cdots (\hbox{ no } k)~.
\label{eq:IV}
\end{eqnarray} 
As shown in Ref.~\cite{paper1}, the original contribution has one
term that is proportional to the lower order matrix element, as well
as a tower of terms that contain only one term that depends on $k$ in
the numerator; in particular this dependence is {\em odd} in $k$, being
either $Q \cdot k$ or $(-2k)_{\mu_q}$, where $Q$ is a combination of the
momenta not involving $k$: $Q= P + \sum_q$. The
remaining $k$ dependence is contained in the overall integration
$\int d^4 k$, the term $b_k$, and the part of the photon propagator,
$D^{t_\mu,t_\nu} (k)$, all of which are symmetric in $(k \leftrightarrow
-k)$. Hence such terms odd in $k$ vanish, leaving behind only the term
that is otherwise $k$ independent and proportional to the lower order
matrix element. Applying the result from Ref.~\cite{paper1}, we have,
\begin{align}
\left\{{\hbox{On-shell}}^{I+II+III+IV} \right\} = &
	S\strut^{t_V,t_1}_{P} \left\{
 - ~\delta_{t_\mu,t_1} ~\delta_{t_\nu,t_1} (2P+l_1)_{\mu_1}
        S\strut^{12}_{P+\sum_1} (2P+2\Sigma_1+l_2)_{\mu_2}
	\cdots \right\}~,
\label{eq:old}
\end{align}
and hence is proportional to the lower order matrix element as well as
the expected thermal delta function factors for both insertions on the
{\em same} leg (see Section \ref{sss:overview}). Note that
in obtaining this result we have dropped terms that are linear in $k$
(such as $Q \cdot k$ and $(-2k_{\mu_q}$) since such odd terms vanish on
integration. In addition, the $k^2$ dependence cancels on inclusion of
the tadpole diagrams as was the case for pure scalar thermal QED.

Now, adding in the extra contributions from all four sets, we see that
the contribution from Set III and the first term from Set I cancels the
total contribution from Set IV (self energy terms); see Eqs.~\ref{eq:I},
\ref{eq:III}, and \ref{eq:IV}. Also, the second term in Set I cancels
the first term of Set II, leaving behind only the second term of Set II,
which adds to the ``on-shell'' contribution, giving the total contribution
from both insertions on the scalar part of the $p$ leg to be,
\begin{eqnarray} \nonumber
{\cal{C}}_{s+2}^{{\rm scalar};\mu, \nu} & \sim & \left\{
{\hbox{On-shell}}^{I+II+III+IV} \right\}
	+ S\strut^{t_V,t_1}_{P+k} ~\delta_{t_\mu,t_V} (2P+2k +l_1)_{\mu_1}  
	\times \nonumber \\
 & &  	S\strut^{t_1,t_2}_{P+\sum_1+k}
 	(2P+2\Sigma_1+2k+l_2)_{\mu_2} 
	\cdots (\hbox{all } k) \times
 	S\strut^{t_s,t_X}_{P+k+\sum_s} ~\delta_{t_\nu,t_X}~,
\end{eqnarray}
where the first term is given in Eq.~\ref{eq:old}. This result for
insertion of both vertices on the scalar part of the $p$ leg, is to be
multiplied with the contribution from the fermionic part of the $p$ leg;
since this has no $k$ dependence, it is simply given by the
corresponding part of the lower order matrix element. We have,
\begin{eqnarray} \nonumber
{\cal{C}}_{s+2}^{{\rm scalar};\mu, \nu} \times
{\cal{C}}_{r}^{{\rm fermion}~p} & \sim & \left\{
	-S\strut^{t_V,t_1}_{P} ~\delta_{t_\mu,t_1} ~\delta_{t_\nu,t_1}
	(2P+l_1)_{\mu_1} S\strut^{12}_{P+\sum_1}
	(2P+2\Sigma_1+l_2)_{\mu_2} \cdots \right. \nonumber \\
 & & 	+ S\strut^{t_V,t_1}_{P+k} ~\delta_{t_\mu,t_V} (2P+2k +l_1)_{\mu_1}  
 	S\strut^{t_1,t_2}_{P+\sum_1+k}
 	(2P+2\Sigma_1+2k+l_2)_{\mu_2} \nonumber \\
 & & 	\left.\cdots S\strut^{t_s,t_X}_{P+k+\sum_s}
	 ~\delta_{t_\nu,t_X} \right\} \times \left\{ \bar{u}_{q'} \Gamma_X
  	S\strut^{t_X,t_r}_{p+\sum_r} ~\delta_{\nu,X} \gamma_r \cdots
  	u_p ~\right\}~.
\label{eq:ppss}
\end{eqnarray}
We now compute the total contribution when both vertices are inserted on the
$p$ leg as the sum of contributions from the three different
types of insertions, viz., both vertices on the fermion part of the $p$
leg, one vertex each on the fermion and scalar part of the $p$ leg, and
both vertices on the scalar part of the $p$ leg (that is, the sum of
Eqs.~\ref{eq:ppff}, Eqs.~\ref{eq:ppsf}, and Eqs.~\ref{eq:ppss}). We find
that the first term of Eq.~\ref{eq:ppsf} cancels against the second term
of Eq.~\ref{eq:ppss}, while the contribution of Eq.~\ref{eq:ppff}
evaluated to zero. The final contribution is therefore,
\begin{eqnarray} \nonumber
{\cal{M}}_{n+1}^{p,p} & \sim & \left\{
	S\strut^{t_V,t_1}_{P} (2P+l_1)_{\mu_1}
  	S\strut^{t_1,t_2}_{P+\sum_1} \cdots
         S\strut^{t_s,t_X}_{P+\sum_s} ~\delta_{t_\mu,t_X} 
  	-S\strut^{t_V,t_1}_{P} ~\delta_{t_\mu,t_1} ~\delta_{t_\nu,t_1}
	(2P+l_1)_{\mu_1} S\strut^{t_1,t_2}_{P+\sum_1} \right. 
	\nonumber \\
 & & 	\left. \cdots S\strut^{t_s,t_X}_{P+\sum_s}
	\right\} 
 	\times \left\{ \bar{u}_{q'} \Gamma_X
  	S\strut^{t_X,t_r}_{p+\sum_r} ~\delta_{\nu,X} \gamma_r \cdots
  	u_p ~\right\}~, \nonumber \\
 & = & \left(\delta_{t_\mu,t_X}  \delta_{t_\nu,t_X}  -
         \delta_{t_\mu,t_1}  \delta_{t_\nu,t_1} \right) \times
 	\left\{S\strut^{t_V,t_1}_{P} (2P+l_1)_{\mu_1}
  	S\strut^{t_1,t_2}_{P+\sum_1} \cdots \right\}~, \nonumber \\
 & = & 0~,
\end{eqnarray}
since $t_X=t_1=1$.

\paragraph{Discussion on the nature of the cancellation}

It is instructive to understand the origin of this result. Note
that it is identical to the result from purely fermionic or purely
scalar QED that contributions from insertions of both vertices of the
virtual $(n+1)^{\rm th}$ $K$ photon on the $p$ leg in all possible ways
vanishes. In the present case, there are three contributions. The one
where both vertices are on the fermionic part of the $p$ leg is zero,
as before. There are two more contributions, one where both vertices
are on the scalar line and one where one vertex is on the scalar line
and the other on the initial fermion line.

The contribution when one vertex each is on the scalar and fermion
lines was expected to give a single contribution that is proportional to
$\delta_{t_\nu,t_X} \delta_{t_\mu,t_X}$; however, due to the off-shell
nature of the scalar line, there is an additional term proportional to
$\delta_{t_\nu,t_X} \delta_{t_\mu,t_V}$ as well.

The contribution when both vertices are on the scalar line are terms
proportional to $\delta_{t_\nu,t_1}\delta_{t_\mu,t_1}$ as expected
from the discussion in Section \ref{sss:overview}; however, there is an
additional term proportional to $\delta_{t_\nu,t_X}\delta_{t_\mu,t_V}$
arising from the off-shell nature of the scalar line. The extra terms in
each of these two contributions cancel each other, leaving behind terms
proportional to $\delta_{t_\nu,t_X} \delta_{t_\mu,t_X}$ and
$\delta_{t_\nu,t_1} \delta_{t_\mu,t_1}$ which cancel against each other
since the thermal factors of all external fields are the same, $t_X=
t_V = t_1 =1$. Hence the cancellation is non-trivial and cannot be
written down just by inspection of the pure fermion or pure scalar case
alone.

\subsubsection{Final matrix element for both vertices of the $K$ photon on
the same leg}

As in the case of thermal QED, we can symmetrise this result since we
could have accounted for wave function renormalisation either on the
$p'$ leg or on the $p$ leg. Putting back the suppressed factors such
as $b_k(p,p)$, the remaining part of the photon propagator, and the
integration over $k$, we have,
\begin{eqnarray} \nonumber
{\cal M}\strut^{p'p', K\gamma}_{n+1} & = &  +\frac{ie^2}{2}
	\int \frac{{\rm d}^4 k}{(2\pi)^4} \, \delta_{t_\mu,t_1} \,
	\delta_{t_\nu,t_1} \, b_k(p',p') \, D^{t_\mu,t_\nu} (k)~{\cal
	M}_n~; \nonumber \\
{\cal M}\strut^{pp, K\gamma}_{n+1} & = &  +\frac{ie^2}{2}
	\int \frac{{\rm d}^4 k}{(2\pi)^4} \, \delta_{t_\mu,t_1} \,
	\delta_{t_\nu,t_1} \, b_k(p,p) \, D^{t_\mu,t_\nu} (k)~{\cal
	M}_n~.
\label{eq:Kgp'p'pp}
\end{eqnarray}
Eq.~\ref{eq:Kgp'p'pp}, along with Eq.~\ref{eq:Kgp'p} shows the
factorisation of all possible insertions of an $(n+1)^{\rm th}$ virtual
$K$ photon into an $n^{\rm th}$ order graph.

\subsubsection{Total matrix element for insertion of a virtual $K$ photon}
Combining the three different kinds of insertions, we have
\begin{align} \nonumber
{\cal{M}}_{n+1}^{K\gamma,{\rm tot}} = & \frac{ie^2}{2} \int
	\frac{{\rm d}^4 k}{(2\pi)^4} \, \left\{ \delta_{t_\mu,t_1} \,
	\delta_{t_\nu,t_1} \, D^{t_\mu,t_\nu} (k) \,
	\left[\strut b_k(p',p') + b_k(p,p) \right] \right. \nonumber \\
	 & +\left. \delta_{t_\mu,t_V} \, \delta_{t_\nu,t_V} \,
	 D^{t_\mu,t_\nu} (k) \,
	\left[\strut -2b_k(p',p) \right] \right\}
	 {\cal{M}}_{n}~, \nonumber \\
	 & \equiv \left[B \right] {\cal{M}}_{n}~,
\label{eq:K}
\end{align}
where the prefactor containing the IR divergence can be expressed as,
\begin{align} \nonumber
B & = \frac{ie^2}{2} \int
	\frac{{\rm d}^4 k}{(2\pi)^4} \, D^{11} (k) \,
	\left[\strut b_k(p',p') - 2 b_k(p',p) + b_k(p,p) \right]~, 
	 \nonumber \\
  & \equiv \frac{ie^2}{2} \int
	\frac{{\rm d}^4 k}{(2\pi)^4} \,
	D^{11} (k) \, \left[\strut J^2(k) \right]~,
\label{eq:B}
\end{align}
where we have used the fact that the thermal types of the hard/external
vertices must be type-1. We see that each term is proportional to the
(11) component of the photon contribution and this is crucial to achieve
the cancellation between virtual and real photon insertions, as we show
below.

\subsection{Insertion of virtual $G$ photons}
\label{ssec:ggamma}

Before we go on to consider the IR divergences from the real photon
insertions, we briefly consider the effect of the insertion of an
$(n+1)^{\rm th}$ $G$ photon to an $n^{\rm th}$ order graph. Here the
earlier results from pure thermal fermionic and scalar QED hold and we
simply outline the proof that such $G$ photon insertions are IR finite.
We note that in contrast to the $T=0$ case addressed by GY where the
leading IR divergences are logarithmic, the presence of the number
operator in the photon propagator (see Eq.~\ref{eq:gammaprop}) increases
the degree of divergence, leading to both linearly divergent as well as
logarithmically sub-divergent IR contributions. The demonstration of IR
finiteness of the leading contribution is straightforward and an
application of the approach of GY; that of the sub-leading contribution
is non-trivial and was dealt with in detail in the companion paper,
Ref.~\cite{paper1}. The proof of IR finiteness for the leading linear
divergence arises from the {\em construction} of $b_k$. Consider a $G$
photon insertion on legs described by momenta $p_f$ and $p_i$ (where
$p_f$ and $p_i$ could be any of $p'$, $p$). Since the hard momentum
$p_f$ ($p_i$) flows through the entire $p_f$ leg ($p_i$ leg), the $G$
photon contribution is proportional to
\begin{align} \nonumber
{\cal{M}}_{n+1}^{G\gamma} & \propto
	\left\{ g_{\mu\nu} - b_k(p_f,p_i) k_\mu k_\nu\right\} \times
	p_f^\mu \, p_i^\nu~, \nonumber \\
 & = 0 + {\cal{O}}(k)~.
\label{eq:Gfermion}
\end{align}
At finite temperature, the terms linear in $k$ are also IR divergent and
hence it is necessary to show the cancellation upto this order. Of
course the $T=0$ part of the result is already IR safe since it is only
logarithmically divergent. Hence it is necessary to consider only the
$T \ne 0$ part of the calculation. A look at the structure of the
propagators (see Eqs.~\ref{eq:gammaprop}, \ref{eq:fprop} and
\ref{eq:sprop} in Appendix \ref{app:abino}) immediately indicates that
all such contributions are on-shell, so $k^2=0$. Then $b_k$ simplifies
to
\begin{equation}
b^{T\ne 0}_k(p_f,p_i) = \frac{p_f \cdot p_i} {p_f \cdot k \; \; p_i \cdot k}~.
\label{eq:bkT}
\end{equation}
With this definition, the $G$ photon insertion turns out to be
\cite{paper1},
\begin{align}
{\cal M}\strut^{G\gamma}_{n+1} \sim & \int d^4 k
	\left[ \frac{i}{k^2+i\epsilon} \delta_{t_\mu,t_\nu} \pm  2 \pi
	\delta(k^2) N(\vert k \vert) D_{t_\mu,t_\nu} \right]
        \left[\strut 0 (p_f \cdot p_i) + 2 (p_f  + 2 p_i) \cdot k \right]
 	\left[\strut {\rm scalar} \right]_{\slashed{\mu}\slashed{\nu}}~,
\label{eq:MGstrsimp}
\end{align}
where the first term is from the inserted photon propagator, and the
slashes on $\mu$ and $\nu$ indicate that the contributions from
insertion of these vertices have been removed and simplified to yield the
terms in the second brackets, and the last term has an expansion in $k$
given by,
\begin{align}
 \left[ {\rm scalar} \right]_{\slashed{\mu}\slashed{\nu}} & \sim 
 \left[ {\cal O}(1) + {\cal O}(k) + {\cal O}(k^2) + \cdots \right]~.
\label{eq:sca}
\end{align}
Now, there are two contributions at ${\cal{O}}(k)$: an ${\cal{O}}(k)$
factor from the second bracket of Eq.~\ref{eq:MGstrsimp} and
${\cal{O}}(1)$ from the third bracket and vice-versa. It can be shown
(see details in Refs.~\cite{Indu} and \cite{paper1}) that both these terms
factor in such a way that all terms odd in $k$ vanish, since the remaining
$k$ dependence arising from the $(n+1)^{\rm th}$ photon propagator, the
integration measure, and $b_k$ are even in $k$. Moreover, for terms which
have no particular symmetry in $k$, the structure of the propagators is
such that, when symmetrised over $k \leftrightarrow -k$, the divergence
is softened by one order, so that there are no remaining contributions
with terms in the numerator that are linear in $k$. Hence the $G$
photon contribution is IR finite with vanishing of both ${\cal{O}}(1)$
and ${\cal{O}}(k)$ terms, the former due to the construction of $b_k$
and the latter due to symmetry arguments.

So far we have simply written down expressions including the fermion and
scalar propagators but have not examined their effects in detail. The
final complexity lies in explcitly including thermal effects in the
fermion and scalar propagators. In the soft limit, the fermion number
operator is well-behaved:
\begin{equation}
N_f(\vert P^0 \vert)  = \frac{1}
	{\exp[\vert P^0 \vert /T] +1} \; 
  \stackrel{P^0\to 0}{\longrightarrow} \; \frac{1}{2}~.
\end{equation}
This is in contrast to the scalar number operator,
\begin{equation}
N_S(\vert P^0 \vert)  = \frac{1}
	{\exp[\vert P^0 \vert /T] -1} \; 
  \stackrel{P^0\to 0}{\longrightarrow} \; \frac{1}{\vert P^0 \vert}~;
\end{equation}
hence it is important to study the soft contribution of thermal scalar
propagators and check whether they spoil the result. This is discussed
in detail in the companion paper where it is shown that the $G$ photon
insertions are IR finite when we consider the entire thermal structure
of the theory, including that of vertex factors and all propagators. The
results hold here as well and the arguments are not reproduced here; the
reader is referred to Ref.~\cite{paper1} for details.

\subsubsection{The final matrix element for virtual $K$ and $G$ photons}

We now specify that the $n^{\rm th}$ order graph contains $n_K$ $K$
photon and $n_G$ $G$ photon insertions, so $n = n_K + n_G$. Since the
insertions are to be symmetrised over all these bosonic contributions,
the total matrix element can be expressed as,
\begin{equation}
\frac{1}{n!}\, {\cal M}_n =
  \sum_{n_K=0}^{n} \frac{1}{n_K!}  \frac{1}{n - n_K!} {\cal
    M}_{{n_G},{n_K}}~.
\end{equation}
Summing over all orders, we get
\begin{align}
\sum_{n=0}^{\infty} \frac{1}{n!} {\cal M}_n & =
\sum_{n=0}^{\infty} \sum_{n_K=0}^{n} \frac{1}{n_K!}
\frac{1}{n - n_K!} {\cal M}_{{n_G},{n_K}}~, \nonumber \\
 & =
\sum_{n_K=0}^\infty \sum_{n_G=0}^{\infty} \frac{1}{n_K!}
\frac{1}{n_G!} {\cal M}_{{n_G},{n_K}}~,
\end{align}
Since the $K$ photon contribution is proportional to the lower
order matrix element we have,
\begin{align} {\cal M}_{{n_G},{n_K}} & =
(B)^{n_K} M_{n_G,0} \equiv (B)^{n_K} M_{n_G}~,
\end{align}
where $B$
as defined in Eq.~\ref{eq:B} is the contribution from each $K$-photon
insertion and can be isolated and factored out, leaving only the
IR finite $G$-photon contribution, ${\cal M}_{n_G}$. Re-sorting and
collecting terms, we obtain the requisite exponential IR divergent
factor:
\begin{align}
\sum_{n=0}^\infty \frac{1}{n!}
{\cal M}_n & = \sum_{n_K=0}^\infty \frac{(B)^{n_K}}{n_K!}
\sum_{n_G=0}^{\infty} \frac{1}{n_G!} {\cal M}_{n_G}~, \nonumber \\
     & = {\rm e}^{B} \sum_{n_G=0}^\infty \frac{1}{n_G!} {\cal
     M}_{n_G}~.
\label{eq:Mvirtual}
\end{align}
Before we compute the cross section, we briefly consider the case of
insertion of real photons.

\subsection{Emission and absorption of real photons}
\label{ssec:realgamma}

There are no additional complications here since only one real photon
vertex can be inserted at a time unlike virtual photons that have two
vertices. Again the insertions must be made in all possible ways to
yield a symmetric result; also, the results of the insertions on the
fermion and scalar lines factor and can be independently considered. The
contribution of the $(n+1)^{\rm th}$ real photon insertion to the cross
section is obtained by squaring the matrix elements and applying the
separation into $\widetilde{K}$ and $\widetilde{G}$ components of the
polarisation sum:
\begin{align}
\sum_{\rm pol} \epsilon^*_\mu (k)\,\epsilon_\nu (k) & = -
	g_{\mu\nu}~, \nonumber \\
 & = -\left\{\strut \left[\strut g_{\mu\nu} -
	\tilde{b}_k(p_f,p_i) k_\mu k_\nu \right] +
	\left[\strut \tilde{b}_k(p_f,p_i) k_\mu k_\nu \right] \right\}~,
	\nonumber \\
 & \equiv -\left\{\left[ \widetilde{G}_{\mu\nu}\right] + \left[
 \widetilde{K}_{\mu\nu}\right]\right\}~.
\end{align}
The key point to note is that all real photons, whether emitted or absorbed,
correspond to photons of thermal type 1; hence the inserted vertex due
to the $(n+1)^{\rm th}$ real photon is of thermal type 1 only. The
significance of obtaining a total virtual photon contribution that is
proportional to $D^{11}$ alone (see Eqs.~\ref{eq:K} and \ref{eq:B})
is now clear; the cancellation between the soft real and virtual photon
contributions can occur only in this case.

There are two points to note: one is the complication that arises since
physical momentum is gained or lost in real photon absorption/emission.
This is taken care of by adjusting the initial bino momentum
appropriately so that $p+q = p'$ still holds. The second is that the
phase space factor for thermal photons is different from the zero
temperature case. We have,
\begin{align}
\d\phi_i & = \frac{\d^4k_i}{(2\pi)^4} \, 2\pi \delta(k_i^2) \,
	\left[ \theta(k_i^0) + N(\vert k^0_i \vert) \right]~;
\label{eq:phase}
\end{align}
again, the thermal modifications include the number operator. Hence,
just as in the virtual photon insertion, the presence of the number
operator worsens the IR divergence to be linear in real photon case as
well. This is essential if the real photon absorption/emission IR
divergent contributions must cancel the corresponding virtual photon
contributions. A straightforward repetition of the earlier analysis for
pure thermal fermions and scalars for the current process gives a
$\widetilde{K}$ contribution that contains the IR divergent part. We
again do not reproduce the calculation here and simply write down the final
contribution to the cross section at the $(n+1)^{\rm th}$ order:
\begin{align} \nonumber
\left \vert {\cal{M}}_{n+1}^{\widetilde{K}\gamma,{\rm tot}} \right \vert^2
	& \propto -e^2 \left[\strut \tilde{b}_k(p,p) -2
	\tilde{b}_k(p',p) + \tilde{b}_k(p',p') \right]~,
		\nonumber \\
   & \equiv -e^2 \widetilde{J}^2(k)~,
\label{eq:ktilde}
\end{align}
where $\tilde{b}_k(p_f,p_i)$ is given by the expression for $b_k$ in
Eq.~\ref{eq:bk} with $k^2=0$, as is consistent for a real photon. The
$\widetilde{G}$ photon contributions can also be proved to be IR finite.
Here, the key observation is that the phase space factor in
Eq.~\ref{eq:phase} is not symmetric under $k \leftrightarrow -k$;
however, the finite temperature part of the phase space is symmetric
under $k \leftrightarrow -k$ since both photon emission into and photon
absorption from the heat bath are allowed. The $T=0$ part of the
$\widetilde{G}$ photon contribution is IR finite by construction of
$\tilde{b}_k$; this is a logarithmic divergence with no left over
subdivergences and hence there is no need for symmetry in the $T=0$
part. Again, it is possible to apply symmetry arguments for the $T \ne 0$
part and show that $\widetilde{G}$ photon insertions are IR finite with
respect to both real photon emission and absorption. Hence the presence of
photon emission and absorption is crucial for showing this. This has also
been observed in Ref.~\cite{Beneke} where the IR finiteness of such a
theory of dark matter was shown to next-to-leading order (NLO). 

\subsection{The total cross section to all orders}
\label{ssec:cross}

We now consider higher order corrections to all orders of the process
$\chi {\cal{F}} \to \chi {\cal{F}}$. All the real photons add a piece to
the energy-momentum delta function that can be included along with
$\widetilde{J}$ for every real $\widetilde{K}$ photon insertion:
\begin{equation}
\widetilde{B}(x) = -e^2 \int \widetilde{J}^2(k_k) \d\phi_k
	\exp\left[ {\pm i\, k_k \cdot x}\right]~,
\label{eq:Btilde}
\end{equation}
where the signs $\pm$ refer to photon emission/absorption respectively.
Hence, the total cross section, including both virtual and real photon
corrections, to all orders, for the process $\chi {\cal{F}} \to \chi
{\cal{F}}$, can be expressed analogously to the result obtained in
Ref.~\cite{paper1} as,
\begin{align} \nonumber
\d\sigma^{\rm tot} & = 
 \begin{aligned}[t]
 \int \d^4x \, e^{-i(p+q-p')\cdot x} \d\phi_{p'} \d\phi_{q'}
          \exp\left[\strut B+B^* \right] \exp \left[\widetilde{B}
	  \right] \times 
  \sum_{n_G=0}^{\infty} \frac{1}{n_G!} \times \\
	  \prod_{j=0}^{n_G} \times
   	\int \d\phi_j e^{\pm i k_j \cdot x}
	  \left[ -G_{\mu\nu} {\cal M}_{n_G}^{\dagger\mu}
	  {\cal M}_{n_G}^\nu \right] ~,
  \end{aligned} \\
 & = \int \d^4x \, e^{-i(p+q-p')\cdot x}\, \d\phi_{p'}  \d\phi_{q'} 
           \exp\left[ B+B^*+\widetilde{B} \right] \,
	   \sigma^{\rm finite} (x)~.
\label{eq:sigmatot}
\end{align}
Here $\sigma^{\rm finite}$ contains both the finite $G$ and $\widetilde{G}$
photon contributions from both virtual and real photons respectively.
The IR finiteness of the cross section can be demonstrated by examining
the soft limit of the virtual and real $K$ photon contributions: we see
that the IR divergent parts of both the virtual and real photon
contributions exponentiate and {\em combine to give an IR finite sum},
as can be seen by studying their behaviour in the soft limit when $k \to
0$::
\begin{align} 
(B+B^*) + \widetilde{B} & = e^2 \int \d\phi_k \left[
	J(k)^2\left\{\strut 1+2N(\vert k^0 \vert)\right\} -
	\widetilde{J}(k)^2 \left\{\strut \left(1+N(\vert k^0 \vert)
	\right)e^{i k\cdot x} +
	N(\vert k^0 \vert)e^{-i k\cdot x}
	\right\} \right] \nonumber \\
 & \stackrel{k \to 0}{\longrightarrow} ~0 + {\cal O}(k^2)~.
\label{eq:finite}
\end{align} 
Notice that the cancellation occurs between virtual and real
contributions upto order ${\cal{O}}(k)$. Furthermore, both real photon
emission and absorption terms were required to achieve this
cancellation. Finally, the cancellation occurred because the
contribution of the thermal type-1 real photons matched that of the
thermal type-1 virtual photons, since there was no contribution from
thermal virtual type-2 photons. Thus, by making use of earlier results
on the IR finiteness of thermal theories of pure fermionic and scalar
QED, we have demonstrated the IR finiteness of a theory of dark matter
particles interacting with charged scalars and fermions to all orders in
the theory. From the structure of the real time formulation, this
implies that the corresponding zero temperature theory is IR finite to
all orders as well.

\section{Discussions and Conclusions}
\label{sec:concl}

The ``WIMP miracle'' is oft-quoted as an argument for the viability
of a generic cold Dark Matter candidate $\chi$. This is because, for
such a particle having interactions with known species with a strength
comparable to the electroweak gauge coupling, the relic abundance
naturally turns out to be of the same order as the observed one. In
particular, if its interactions with the SM fermions ($\mathcal{F}$)
are mediated by charged scalars with the relevant Yukawa couplings 
being of the aforementioned order, then it freezes out at a temperature
$T_{\rm freeze} \sim m_\chi/20$. As is evident, it is the details of
the model that would determine the exact value of $T_{\rm freeze}$ and,
hence, that of the relic energy density. The rather precise measurements
of the latter by {\sc wmap} and {\sc planck} have, thus, imposed severe
constraints on the parameter space of models, including popular ones
such as the MSSM or those with extended gauge symmetry. Furthermore,
the parameter space favoured by relic abundance considerations militates
against the continual non-observation at satellite-based direct detection
experiments, or even the large hadron collider.

Given this, it is extremely important to reconsider if, in the
calculation of relic abundances, important corrections have been
overlooked.  While some efforts have been made to this end, most of
them were done at zero temperature. This, clearly, is not enough as
the DM is touted to have decoupled (and the relic abundance
established) at high enough temperatures for finite temperature
effects to be of relevance in the calculation of cross
sections. Indeed, Ref.~\cite{Beneke} showed the isolation and
cancellation of infra-red divergences to NLO and calculated the
corresponding infra-red finite cross sections in the thermal
theory. Our aim was more ambitious in that we wanted to establish the
all-order infrared finiteness of such theories. The infrared finitenes
to all orders of a thermal field theory of pure charged fermions was
already established \cite{Indu}. Clearly, in order to establish the
infrared finiteness of the thermal theory of dark matter, a
corresponding result for pure charged scalars in a heat bath was required.
Hence, in a companion paper \cite{paper1}, the infrared finiteness of
such a thermal field theory was proven to all orders.

In this second paper, these results on thermal fermionic and thermal
scalar QED were applied to the theory of dark matter interactions. First
and foremost, rather than get embroiled in the details of a specific
model, we began by distilling the essence of such models for a fermionic
dark matter particle, two of which can annihilate to a pair of SM fermions
via the exchange of a charged scalar, as shown in Fig.~\ref{fig:dm}. This
may, at first glance, seem to be only a particular formulation of the DM
with other extensions of the SM allowing for a fermionic DM annihilating
to charged scalars through a fermion exchange. Even more different
would look a theory of scalar DM annhilating to SM fermions through the
exchange of charged fermions. However, a moment's reflection would assure
one that all such cases can be reinterpreted in terms of the basic block
that we consider here. Indeed as we explicitly point out in this paper,
the analysis is ``blind'' to the precise structure of the hard process at
vertices $V$ and $X$ where the dark matter particle interacts.  Nowhere in
our analysis did we use the actual structure of this vertex. Only the
vertices and propagators arising from additional photon (real or virtual)
insertions were germane to the issue. Thus, even if the hard vertex had
been a different one (say, one corresponding to a general coupling $J(x)
\phi(x)$, where $J(x)$ is an arbitrary current), the same analysis would
have gone through. A particular example of such a coupling would be the
Yukawa theory, {\em viz.}, $\bar {\mathcal{F}}_1 \, (a_1 + a_2 \gamma_5)
\, {\mathcal{F}}_2$, where ${\mathcal{F}}_i$ are (potentially different)
fermions and $a_{1,2}$ arbitrary couplings.

In summary, real-time formulations of thermal field theories lead to
field doubling with external observable particles corresponding to
type-1 fields and propagators allowing for transformation between type-1
and type-2 fields. The structure of the propagator then takes a $2
\times 2$ form and vertex factors also acquire thermal dependence.
Similarly, the phase space for real photon emission into, and absorption from
the heat bath, is also modified. In both the phase space and the
propagator elements, the thermal modifications include the number
operator that worsens the degree of divergence in the soft limit from
logarithmic to linear. 

Our key findings here were that (1) the structure of the thermal vertex
was such that the IR-divergent part that factors out in the {\em virtual}
part of the matrix element is proportional to $D^{11}$, that is, the
(11) matrix element of the thermal photon propagator, (2) that the
contribution from {\em real} photon insertions have vertex factors that are
of thermal type-1 alone, thus enabling a cancellation between real and
virtual contributions, and (3) that both {\em emission and absorption
of real photons} with respect to the heat bath was required to achieve
the cancellation of IR divergences.

Both the virtual and real photon contributions have a $T=0$ part that
corresponds to the zero temperature field theory. Hence this computation
establishes the infrared finiteness of the zero temperature theory as
well. Also, as seen from Eq.~\ref{eq:finite}, the finite temperature terms
that are dependent on $N(\vert k^0 \vert)$ cancel only in the
soft limit when the exponential terms from the real photon contribution
reduce to $\exp(\pm ik \cdot x) \to 1$. However, only soft real photons
are included in the establishment of the cancellation; in other words,
for any resolvable photon energy {\em above} the IR cutoff, the signature
of the thermal bath would be discernible.

It was pointed out in Ref.~\cite{BenekeOPE} where the NLO result
obtained in Ref.~\cite{Beneke} was re-established using the operator
product expansion (OPE) technique, that in an OPE approach the relevant
operators are such that the IR divergences (both soft and collinear)
do not appear in the corresponding coefficient functions and are hence
IR finite. While the present work may lack the computational elegance of
such an approach, it offers important insights into the nature of the 3-
and 4-point vertices in scalar-photon interactions (as detailed both in
Paper I and this paper), and the explicit role of the heat bath in both
emitting and absorbing photons.

Furthermore, the formalism delineated here lends itself more easily to
cases where, in scattering (or annihilations), the {\em in}-state has more
than two particles, a situation that is not uncommon to theories of DM,
especially for (but not limited to) those with, say a $Z_3$ symmetry
(unlike the more common $Z_2$ symmetry) protecting the stability of
the DM.  Having established the IR finiteness of the complete theory,
we can now go ahead and compute various (finite) cross sections of
interest, as was done in Ref.~\cite{Beneke}.  Several techniques for
calculating the finite remainder exist, including renormalisation group
methods, the use of $G$ photon insertions described here {\rm etc.}.
Being model-specific, such calculations are beyond the scope of this work.

\paragraph{Acknowledgements}

We thank M. Beneke for bringing the
results of Ref.~\cite{BenekeOPE} to our attention after reading Paper I.

\appendix
\renewcommand{\theequation}{\thesection.\arabic{equation}}

\section{Feynman rules for field theories at finite temperature}
\label{app:abino}

\setcounter{equation}{0}

The Lagrangian corresponding to a thermal field theory of binos
interacting with charged scalars and fermions can be written down,
starting from the zero-temperature Lagrangian given in Eq.~\ref{eq:L}
\cite{thermal}. This gives both the propagators as well as vertex
factors relevant to the thermal theory.

\begin{figure}[htp]
\centering
\includegraphics[width=0.7\textwidth]{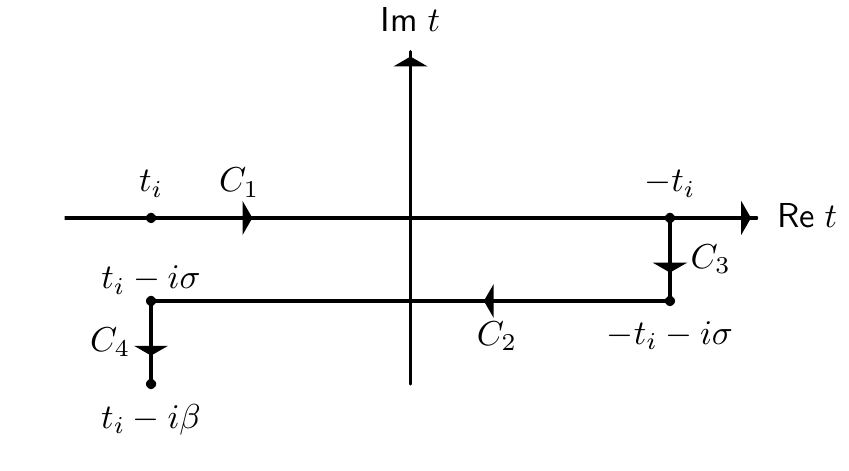}
\caption{\em Standard time path for real time formulation of thermal field
theories. The type-1 and type-2 thermal fields ``live'' on the $C_1$ and
$C_2$ paths of the contour.}
\label{fig:timepath}
\end{figure}

In the path integral formulation of the real time formulation of a
thermal field theory, the generating functional $Z_C(\beta; j)$ (where
$Z_C(\beta;0)$ is the usual partition function), is used to define
averages/expectation values of time ordered products where the time
ordered path $C$ is in a complex time plane from an initial time, $t_i$
to a final time, $t_i - i \beta$, where $\beta$ is the inverse temperature
of the heat bath, $\beta = 1/T$; see Fig.~\ref{fig:timepath}. The
consequence is that these thermal fields satisfy the periodic
boundary conditions,
\begin{equation*}
\varphi(t_0) = \pm \varphi(t_0 - i \beta)~,
\end{equation*}
where $\pm 1$ correspond to boson and fermion fields respectively.

The photon propagator in the Feynman gauge is given by,
\begin{align} \nonumber
i {\cal{D}}_{\mu\nu}^{t_at_b}(k) & = {-g_{\mu\nu}}i{\cal D}^{t_at_b}(k)~,
\nonumber \\
i {\cal D}^{t_at_b}(k) & = \left(\begin{array}{cc} \Delta(k) & 0 \\ 0 &
\Delta^*(k) \end{array} \right) + 2\pi \delta(k^2) N(\vert k^0
\vert ) 
\left(\begin{array}{cc} 1 & e^{\vert k^0 \vert /(2T)} \\
e^{\vert k^0 \vert /(2T)} & 1 
\end{array} \right)~,
\label{eq:gammaprop}
\end{align}
where $\Delta(k) = i/(k^2 + i \epsilon)$, and $t_a, t_b (=1,2)$ refer to the
field's thermal type. The thermal fermion and scalar propagators at zero
chemical potential are given by
\begin{align} \nonumber
i {\cal{S}}_{\rm fermion}^{t_at_b} (p,m) & = \left( \begin{array}{cc} S &
0 \\ 0 & S^* \end{array} \right) - 2 \pi S' \delta(p^2-m^2) N_f(\vert
p^0 \vert) 
\left(\begin{array}{cc} 1 & \epsilon(p_0) e^{\vert p^0 \vert /(2T)} \\
-\epsilon(p_0) e^{\vert p^0 \vert /(2T)} & 1 
\end{array} \right)~, \\
 & \equiv (\slashed{p}+m) \left(\begin{array}{cc} F_p^{-1} & G_p^{-1} \\
                   -G_p^{-1} & F_p^{*-1} \end{array} \right)~,
\label{eq:fprop}
\end{align}
where $S = i/(\slashed{p} -m+i\epsilon)$, and $S' = (\slashed{p} +m)$;
hence the fermion propagator is proportional to $(\slashed{p}+m)$. 
\begin{align}
i {\cal{S}}_{\rm scalar}^{t_at_b} (p,m) & = \left( \begin{array}{cc}
\Delta(p) & 0 \\ 0 & \Delta^*(p) \end{array} \right) +
2 \pi \delta(p^2-m^2) N(\vert p^0 \vert) 
\left(\begin{array}{cc} 1 & e^{\vert p^0 \vert /(2T)} \\
e^{\vert p^0 \vert /(2T)} & 1 
\end{array} \right)~,
\label{eq:sprop}
\end{align}
where $\Delta(p) = i/(p^2-m^2+i\epsilon)$. The first term in each case
corresponds to the $T=0$ part and the second to the finite temperature
piece; note that the latter contributes on mass-shell only.

\begin{figure}[htp]
\centering
\includegraphics[width=\textwidth]{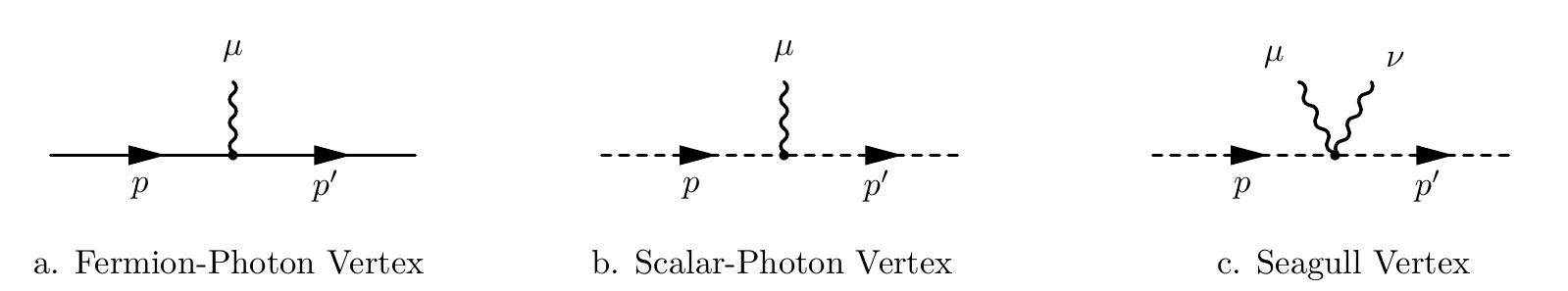}
\caption{\em Allowed vertices for fermion--photon scalar--photon interactions.}
\label{fig:feynman}
\end{figure}

The fermion--photon vertex factor is $(-ie\gamma_\mu)(-1)^{t_\mu+1}$
where $t_\mu=1,2$ for the type-1 and type-2 vertices. The corresponding
scalar--photon 3-vertex factor is $[-ie(p_\mu + p'_\mu)](-1)^{t_\mu+1}$
where $p_\mu$ ($p'_\mu$) is the 4-momentum of the scalar entering
(leaving) the vertex. In addition, there is a 2-scalar--2-photon
{\em seagull} vertex (see Fig.~\ref{fig:feynman}) with factor
$[+2ie^2 g_{\mu\nu}](-1)^{t_\mu+1}$ and a {\em tadpole} diagram
which is also a 2-scalar--2-photon (loop) vertex with factor
$[+ie^2g_{\mu\nu}](-1)^{t_\mu+1}$; note the absence of the symmetry
factor 2 in the latter. All fields at a vertex are of the same type, with
an overall sign between physical (type 1) and ghost (type 2) vertices.

The bino-scalar-fermion vertex factor at a vertex $V$ is denoted as
$\Gamma_V$; for details on Feynman rules for Majorana particles at zero
temperature, see Ref.~\cite{Denner}. These rules apply to the type-1
thermal bino vertex; an overall negative sign applies as usual to the
type-2 bino vertex; again all fields at a vertex are of the same type.

\section{Useful identities at finite temperature}
\label{app:fidentities}

\setcounter{equation}{0}

Various identities useful for fermions are given in Ref.~\cite{Indu}
and are reproduced here for completeness. The corresponding identities for
scalar fields given in Ref.~\cite{paper1} are also listed below.

\begin{enumerate}

\item {\bf The propagators}:
\begin{align} \nonumber
(\slashed{p}-m) ~ i{\cal S}_{\rm fermion}^{t_at_b}(p,m) & = i (-1)^{t_a+1}
\delta_{t_a,t_b}~,
\nonumber \\
(p^2-m^2) ~ i{\cal S}_{\rm scalar~}^{t_at_b}(p,m) & = i (-1)^{t_a+1}
\delta_{t_a,t_b}~.
\end{align}
Henceforth we shall suppress the subscript of scalar or fermion since
the context will be clear. We shall also use the compressed notation,
$i{\cal S}^{t_at_b}(p,m) \equiv i S^{ab}_p$.

\item {\bf The generalised Feynman identities}:
Consider an $n^{\rm th}$ order graph with $s$ vertices labelled $u$ to 1
from the hard vertex $V$ to the right (see Fig.~\ref{fig:darkvertex}).
We now insert
the $\mu$ vertex of the $(n+1)^{\rm th}$ $K$ photon with momentum $k$
between vertices $q+1$ and $q$ on the $p'$ fermion leg. Here the vertex
label codes for both the momentum and the thermal type: the momentum
$p'+\sum_{i=1}^{q}l_i$ flows to the left of the vertex $q$ on the $p'$
fermion leg. The photon at this vertex has momentum $l_q$, with Lorentz
index $\mu_q$, and thermal type-index $t_q$. Denoting $(p'+\sum_{i=1}^q
l_i)$ as $p'+\sum_q$, we have,
\begin{align}
S\strut^{q\mu}_{p'+\sum\limits_q} \, \slashed{k} \,
	S\strut^{\mu,q+1}_{p'+\sum\limits_q+k} & = i
	(-1)^{t_\mu+1} \left[S\strut^{q,q+1}_{p'+\sum\limits_q}
	\delta_{t_\mu,t_q+1} - S\strut^{q,q+1}_{p'+\sum\limits_q+k}
	\delta_{t_\mu,t_q} \right]~.
\label{eq:fid}
\end{align}
Similarly, for the insertion of a virtual $K$ photon at the vertex
$\mu$ on the $p'$ scalar leg, we have,
\begin{align}
S\strut^{q\mu}_{p'+\sum\limits_q} \left[(2p'+2\Sigma_q+k)\cdot k\right]
	S\strut^{\mu,q+1}_{p'+\sum\limits_q+k}
	& = i (-1)^{t_\mu+1} \left[ S\strut^{q,q+1}_{p'+\sum\limits_q}
	~\delta_{t_\mu,t_q+1} - S\strut^{q,q+1}_{p'+\sum\limits_q+k}
	~\delta_{t_\mu,t_q} \right]~.
\label{eq:sid}
\end{align}
If the photon vertex is inserted to the right of the vertex labelled
'1' on the scalar leg with momentum $p'$, we have,
\begin{align} \nonumber
[(2p'+k)\cdot k] S\strut^{\mu,1}_{p'+k} & = (-1)^{t_\mu+1}
\delta_{t_\mu,t_1}~,
\end{align}
since $p'^2 = m^2$. Similar relations hold for the insertion of
of the virtual $K$ photon at a vertex $\nu$ on the scalar $p$ leg since
$p^2 = m^2$ as well.

\end{enumerate}

\end{document}